\documentclass[reprint, amsmath, amssymb, superscriptaddress]{revtex4-2}

\usepackage{graphicx}
\usepackage{natbib}
\usepackage{amsmath}
\usepackage{caption}
\usepackage{float}
\usepackage{dcolumn}
\usepackage{bm}
\usepackage[colorlinks=true,citecolor=blue,linkcolor=blue,urlcolor=blue]
{hyperref}
\usepackage{subfig}
\usepackage{multirow}

\begin{document}
	
	\title{Particle motion around traversable wormholes: \\ Possibility of closed timelike geodesics}
	
	
	\author{Ayanendu Dutta}
	\email{ayanendudutta@gmail.com}
	\affiliation{Department of Physics, Jadavpur University, Kolkata-700032, INDIA}
	
	\author{Dhritimalya Roy}
	\email{rdhritimalya@gmail.com}
	\affiliation{Department of Physics, Jadavpur University, Kolkata-700032, INDIA}
	
	\author{Subenoy Chakraborty}
	\email{schakraborty.math@gmail.com}
	\affiliation{Department of Mathematics, Jadavpur University, Kolkata-700032, INDIA}

	
	\begin{abstract}
	The present work investigates the general wormhole solution in Einstein gravity with an exponential shape function around an ultrastatic and a finite redshift geometry. The geodesic motion around the wormholes is studied in which the deflection angle of the orbiting photon sphere is found to be negative after a certain region, indicating the presence of repulsive effect of gravity in both the ultrastatic and finite redshift wormholes. Various unbounded and bounded timelike trajectories are presented on the wormhole embedding diagrams, in which some of the bound orbits involve intersection points that may lead to causality violating geodesics. Another class of closed timelike geodesics are obtained in the unstable circular trajectory that appeared at the wormhole throat. Finally, the trajectories are classified in terms of the family of CTG orbits.
	\end{abstract}
	
	\keywords{Traversable wormhole; Particle trajectory; Closed timelike geodesics}
	
	\maketitle

\section{Introduction}
In General Relativity, the Einstein's Field equation offers a number of interesting solutions, wormholes being one of them. Wormholes may be considered as bridges or gateways between two points of the same asymptotically flat universe or between two different universes. The concept of this type of interconnecting bridges between space-times was first put forward by Einstein and Rosen \cite{einstein1935} in the year 1935. Their exact solution is the example of the first ever wormhole and was called The Einstein-Rosen bridge. The word 'Wormhole' which is so popular among physicists today was first coined by Misner and Wheeler \cite{misner1957}. Since then a lot of attention has been paid to their geometries and physical effects, traversability being the front runner. Traversability of wormholes were first examined by Morris and Thorne \cite{morris1988, morris1988prl}. Their groundbreaking work which serves as the pillar of wormhole physics provided the first ever solution of the Einstein Field Equation for a traversable wormhole. The stress-energy components obtained for a traversable wormhole always violates the null energy condition  \cite{morris1988, visser-LW, flamm1916, wheeler1955, wheeler1962, hawking1988, geroch1967}. As it is the weakest of all classical energy conditions, it signifies all other energy conditions are also violated. In general, the energy conditions are obeyed by all classical matter, hence the construction of traversable wormhole requires the presence of exotic matter, for instance, the Ellis wormholes of General Relativity \cite{ellis1973, ellis1979, bronnikov1973, kashargin2008_1, kashargin2008_2, kleihaus2014, chew2016}. On the other hand, if modified theories of gravity is considered, the requirement of exotic matter to obtain traversable wormhole is highly reduced, for example see \cite{fukutaka1989, hochberg1990, ghoroku1992, furey2005, bronnikov2010, kanti2011, kanti2012, harko2013}.
	
In the context of exact solution of wormhole geometries and corresponding energy conditions in Einstein gravity and in different modified gravity theories which may yield various interesting results, readers are referred to articles on wormhole geometries in modified gravity, such as in $ f(R) $ gravity \cite{lobo2009, shamir2020,mishra2021}, in $ f(R,T) $ gravity \cite{moraes2017, yousaf2017, elizalde2018, sharif2019}, non-local gravity \cite{Capozziello:2022zoz}, and other theories \cite{sc1,sc2,sc3,sc4,sc5,sc6,sc7,sc8,sc9,sc10,sc11,sc12,sc13,sc14,sc15,sc16,sc17,epl_paper}.

While discussing the traversability, one of the important topics of investigation is the motion of particles in wormhole space-time, especially in traversable wormholes. It is a matter of interest to check whether the geodesics can pass through the throat of the wormhole. Several works have been carried out on the account for different types of wormholes and shape functions. Motion of test particles in Schwarzschild-like wormhole spacetime have been studied by Cataldo \textit{et al}. \cite{cataldo2017}. Their results show that particles moving along the radial geodesics reach the throat with zero tidal velocity in finite time, while the particle velocity reaches maximum at infinity if it travels along a radially outward geodesic, where as for non-radial geodesics, the particles may cross the throat with some restrictions. A detailed study of the geodesic structure for three possible wormholes configuration: Reissner–Nordstr\"om-like wormhole, Schwarzschild-like wormhole and Minkowski-like wormhole was done by Olmo \textit{et al.} \cite{olmo2015}. They concluded that it is possible to have geodesically complete paths for all these wormhole space-times. Geodesic structure in non-static rotating traversable `Teo' wormhole was studied by Chakraborty and Pradhan \cite{chakraborty2017}. Ellis \cite{ellis1973} also constructed a static, spherically symmetric, geodesically complete, horizon less spacetime manifold with a topological hole (drainhole) at its center by coupling the geometry of Schwarzschild spacetime to a scalar field. A detailed general study on null and timelike geodesics were investigated by Mishra and Chakraborty for dynamic spherically symmetric wormhole, rotating wormhole and for Morris-Thorne wormhole with a special attention to Ellis wormhole as well \cite{mishra2018}. They examined the photon spheres and the angle between radial and tangential vectors on the photon trajectory using Rindler and Ishak method. Taylor investigated the traversability conditions for geodesics in a class of wormhole geometry in terms of the radius and shape of the wormhole throat \cite{taylor2014}. The geodesics can be traversed through the throat, reflected by the potential of a wormhole or trapped in the unstable bound orbit at the throat. Here, the geodesic trajectories are visualized on the wormhole embedding diagram by numerically solving the geodesic equations. Recently, geodesic motion around a class of conformally coupled scalar field traversable wormhole (constructed by Barcelo and Visser \cite{barcelo1999}) in Einstein gravity was studied by Willenborg \textit{et al.} \cite{willenborg2018}. They solved the geodesic equations analytically and discussed the possibility of various orbit types and parameter dependence. For recent studies on geodesic motion and photon sphere at the wormhole throat, readers may refer to \cite{godani2,godani3,godani4,godani5}.

In this present work, motivated by the above studies, we intend to investigate the geodesic motion of test particles around a traversable wormhole while also focusing on the possibility of closed timelike geodesics within their orbits. In section \ref{solution}, we have reviewed the wormhole model in Einstein's gravity, and further calculation is added for the field equations with an exponential shape function in a specified redshift geometry. At the same time, respective null and weak energy conditions are studied for the geometry in section \ref{en-con}. Section \ref{geometry} is dedicated to discuss the wormhole geometry and geodesic equations where the deflection angle of photons, the circular timelike geodesics are also obtained. In section \ref{orbits}, we exhibited various geodesic trajectories on the wormhole embedding diagram, prior to the classification of various trajectories. We ended the study with concluding remarks in section \ref{conclusion}.

\section{Mathematical Model of the Wormhole Solution}\label{solution}
In this section, we present a brief review of the mathematical background for construction of a traversable wormhole solution prior to the analysis of particle dynamics around the geometry. Among various possibilities, we undergo the restricted choices of Morris-Thorne type shape function and redshift function while discussing their effect on the geodesic motion.

The static spherically symmetric Morris-Thorne wormhole is given by the metric,
\begin{equation}
	ds^2= - e^{2 \Phi(r)} dt^2+(1-b(r)/r)^{-1} dr^2+r^2 d\Omega^2.
	\label{math01}
\end{equation}
where, $ d\Omega^2=d\theta^2+sin^2\theta d\phi^2 $ is the metric on the two sphere $ \mathbb{S}^2 $, and $ b(r) $, $ e^{\Phi(r)} $ denotes the shape function and redshift function of the wormhole, respectively. Here, the two angular coordinates $ \theta $ and $ \phi $ respectively range from $ 0~\text{to}~\pi $ and $ 0~\text{to}~2\pi $. It is pointed out that the redshift function must be finite everywhere to ensure that the spacetime is free from singularities and horizons. On the other hand, the shape function and the redshift function must hold the following properties:
\begin{enumerate}
	\item $ b(b_0)=b_0 $, as the radial coordinate has a minimum at the wormhole throat ($ b_0 $ be the throat radius). Thus, $ r $ covers from $ b_0~\text{to}~\infty $ in the radial direction.
	\item $\frac{b(r)}{r} \le 1$ for $ r>b_0 $, so that the radial component of the metric $ g_{rr}<0 $ (equality sign only hold at the throat),
	\item $\frac{b(r)}{r} \rightarrow 0 $ and $\Phi(r) \rightarrow 0 $ as $ r \rightarrow \infty $, to ensure the asymptotically flat geometry,
	\item $ \frac{b(r)-r b'(r)}{b(r)^2} >0 $ for $ r>b_0 $, which is nothing but the \textit{flaring-out} condition, and
	\item $ b'(r) \le 1 $ at $ r=b_0 $.
\end{enumerate}

Further, by neglecting any source of matter coupling, the energy-momentum component of any anisotropic fluid is given by $ T^{\nu}_{\mu}=diag(-\rho,~p_r,~p_t,~p_t) $. Here, $ \rho,~p_r,~\text{and}~p_t $ respectively denote the energy density, radial and tangential pressures of the fluid. Now, simple calculations of Einstein field equation for the geometry given by \eqref{math01} yield,
\begin{eqnarray}
	\rho &=& -\frac{b'}{8\pi r^2},
	\label{math02} \\
	p_r &=& -\frac{b}{8\pi r^3} + \left(1-\frac{b}{r}\right) \frac{\Phi'}{4\pi r},
	\label{math03} \\
	p_t &=& \frac{1}{8\pi} \left(1-\frac{b}{r}\right) \Big(\Phi'' +(\Phi')^2 -\frac{rb'-b}{2r(r-b)}\Phi' 
	\nonumber
	\\ &&+\frac{\Phi'}{r} -\frac{rb'-b}{2r^2 (r-b)}\Big),
	\label{math04}
\end{eqnarray}
where the prime denotes differentiation with respect to $ r $. We considered the natural units in the calculation for which $ G=c=1 $.

At this point, we are going to analyze the characteristics and possible matter content of the wormhole model. For that purpose, we shall consider a special type of shape function of the wormhole, i.e. the exponential shape function, which is given by
\begin{eqnarray}
	b(r)=r e^{-(r-b_0)}.
	\label{shape-eq}
\end{eqnarray}
The characteristics and properties of the shape function have already been discussed in numerous studies \cite{godani2020,samanta2020,tangphati2020} establishing that it nicely fits with the aforementioned properties. In \cite{samanta2020}, the significance of the shape function and the tidal force experienced by a wormhole traveler in this particular case are discussed in detail. In this article, the choice of the exponential shape function is purely arbitrary, however, the results (particularly the geodesic movements of test particles) coming out of this shape also held by other usual shape functions and wormholes, e.g. Ellis wormhole and wormhole supported by conformally coupled massless scalar field. It will be briefly discussed in the following sections.

On the other hand, to discuss the wormhole properties for the shape function, first a zero-tidal force wormhole is considered. The zero-tidal force wormhole, also known as the ultrastatic wormhole, always has a particular point of interest. Here, $ \Phi(r)=0 $ i.e. in a gravitational acceleration free frame, if a particle is dropped from rest, remains at rest \cite{morris1988,morris1988prl,cataldo2017}. Later on, we introduce a tidal force on the geometry, according to the redshift function whose expression is provided by,
\begin{eqnarray}
	\Phi(r)=\frac{\gamma}{2r},
	\label{red-eq}
\end{eqnarray}
where $ \gamma $ is an arbitrary constant. $ \gamma $ can possess any value but zero that keeps the redshift finite. Now, substituting the shape and redshift function into the field equations \eqref{math02}, \eqref{math03}, \eqref{math04}, the $ \rho,~p_r,~\text{and}~p_t $ takes the form,
\begin{eqnarray}
	\rho &=& -\left(\frac{1-r}{8\pi r^2}\right) e^{(b_0-r)},
	\label{math05} \\
	p_r &=& \frac{1}{8\pi r^3}\left(-\gamma+(\gamma-r) e^{(b_0-r)}\right),
	\label{math06} \\
	p_t &=& \frac{\gamma}{16\pi r^3} \left(1+\frac{\gamma}{2r}\right) \left(1-e^{(b_0-r)}\right)+\frac{e^{(b_0-r)}}{16\pi r} \left(1-\frac{\gamma}{2r}\right).
	\label{math07}
\end{eqnarray}

Notice that, for $ \gamma=0 $, the zero-tidal force wormhole is recovered. Hence, in this literature, the study of zero-tidal force wormholes is carried out in the form of $ \gamma=0 $. Subsequently, a rather convenient choice of $ \gamma $ i.e. $ \gamma=1 $ is considered for the analysis of non-zero tidal force throughout the study. 

\section{Energy Conditions}\label{en-con}
For the construction of a traversable wormhole, it is worth noting that the wormhole throat must be threaded by a matter that necessarily violates the null and weak energy conditions (NEC and WEC), called as exotic matter or phantom matter.

In the point of view of GR, the inequality $ T_{\mu \nu} X^\mu X^\nu \ge 0 $ corresponding to the energy-momentum tensor for any timelike vector ($ X \in $ any tangent space on the 4-D space-time), must be followed where the local energy density for an observer is denoted by $ T_{\mu \nu} V^\mu V^\nu $ for unit tangent vector $ V $ on the worldline of the observer. This corresponds to the NEC, which in principle pressure form is written as $ \rho+p_i \ge 0, \forall~ i $. The WEC admits the NEC ($ T_{\mu \nu} X^\mu X^\nu \ge 0 $, for any timelike vector $ X^\mu $), along with the claim that for any timelike observer, the local energy density ($ \rho $) must be positive, i.e. in principle pressure form WEC describes $ \rho \ge 0 $ and $ \rho+p_i \ge 0, \forall~ i $.

\begin{figure*}[t]
	\centering
	\subfloat[\label{plot1a}]{{\includegraphics[scale=0.55]{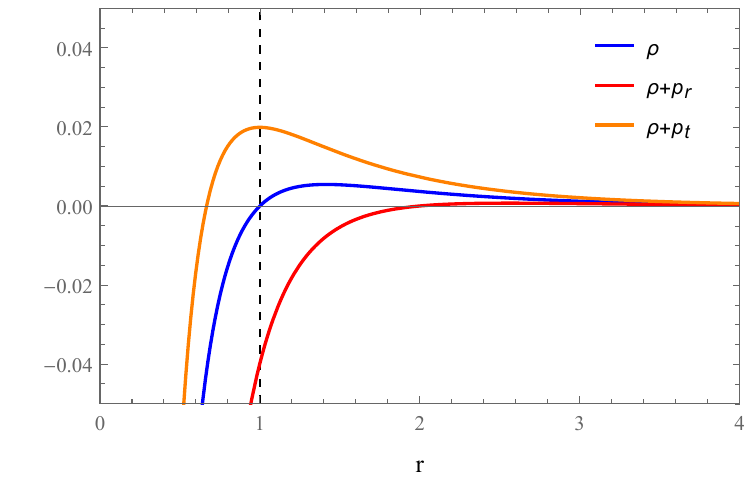}}}\qquad
	\subfloat[\label{plot1b}]{{\includegraphics[scale=0.55]{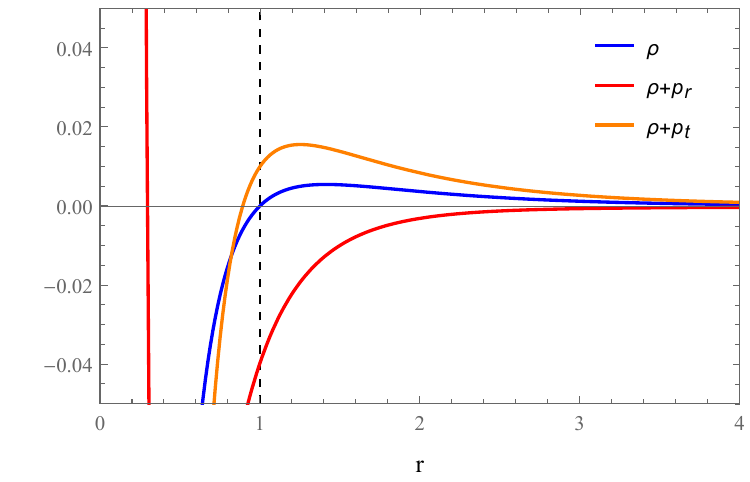}}}
	\caption{The variation of $ \rho,~(\rho+p_r),~(\rho+p_t) $ as a function of radial parameter $ r $, for checking the NEC and WEC. The throat radius $ b_0=1 $ is considered for (a) zero-tidal force wormhole (i.e. $ \gamma=0 $), (b) non-zero tidal force wormhole with $ \gamma=1 $.}
	\label{en_con-plot}
\end{figure*}

Now, for the current study, the calculation of $ \rho,~(\rho+p_r),~\text{and}~(\rho+p_t) $ is rather straightforward from Eq. \eqref{math05}, \eqref{math06} and \eqref{math07}. Plots of the three energy condition components for zero-tidal force (i.e. $ \gamma=0 $) and non-zero tidal force (i.e. $ \gamma=1 $) wormhole are exhibited in Fig. \ref{en_con-plot}. Further, the transition points from positive to negative and vice versa have been listed in Table \ref{en_con-table}. However, it must be noted that the points having a value less than the throat radius are not in the scope of the paper. 

It is observed that irrespective of the tidal force, $ \rho \ge 0 $ in $ r \in [\ b_0, \infty )\ $, the equality only holds at the throat. Alongside, $ (\rho+p_t) $ is satisfied for both the cases. Still, the NEC and thus WEC is violated due to the violation of $ (\rho+p_r) $. For $ \gamma=0 $, $ (\rho+p_r) $ is violated for $ r \in [b_0,2] $, thus supporting the presence of exotic matter. But, for $ r \in [\ 2,\infty)\ $, NEC and WEC are satisfied. In a non-zero tidal force wormhole, i.e. $ \gamma=1 $, $ (\rho+p_r) $ is completely violated, making the presence of exotic matter possible throughout the geometry. Hence, the results support the well-established theoretical foundation for construction of a traversable wormhole.

\begin{table*}[t]
	\centering
	\begin{tabular}{ p {2.5 cm} p {3 cm} p {5 cm} }
		\hline
		\hline
		~~$ \gamma $		&	Terms				&	Result \\
		\hline
		~~$ \gamma=0 $		&	$ \rho $			&	$ \ge 0 $ for $ r \in [\ 1,\infty )\ $\\
							&	$ \rho+p_r $		&	$ \ge 0 $ for $ r \in [\ 2,\infty )\ $\\
							&	$ \rho+p_t $		&	$ \ge 0 $ for $ r \in [\ 0.67,\infty )\ $\\
		\hline
		~~$ \gamma=1 $		&	$ \rho $			&	$ \ge 0 $ for $ r \in [\ 1,\infty )\ $\\
							&	$ \rho+p_r $		&	Always $ < 0 $\\
							&	$ \rho+p_t $		&	$ \ge 0 $ for $ r \in [\ 0.89,\infty )\ $\\
		\hline
		\hline
	\end{tabular}
	\caption{Numerical results for the possible regions where respective energy condition components are satisfied.}
	\label{en_con-table}
\end{table*}

\section{The Geometry and Geodesics}\label{geometry}
Prior to the discussions on geodesics and particle movement in traversable wormholes, a few relevant ideas must be pointed out first.	

The shape of the wormhole can always be visualized for constant time slices in the equatorial plane. We embed the geometry into a higher dimensional ordinary 3D Euclidean space given by the metric
\begin{equation}\label{geo01}
	ds^2= dz^2 + dr^2+ r^2 d\phi^2.
\end{equation}
We can now  show that the embedding function is
\begin{equation}\label{geo02}
	\frac{dz}{dr}=\pm \left( \frac{r}{b(r)}-1 \right)^{-1/2}.
\end{equation}
The radial coordinate ranges from spatial infinity $ r=\infty $ to the throat radius $ r_{min}=b_0 $. The embedding function $ z(r) $ flares out in the throat region, where the \textit{flaring-out} condition is obtained by inverting $ z(r) $ as
\begin{equation}\label{geo03}
	\frac{d^2r}{dz^2}=\frac{b(r)-r b'(r)}{2 b(r)^2} > 0~~~~~~\text{for}~r\ge b_0.
\end{equation}
Further, referring to Eq. \eqref{math01}, it is noted that radial coordinate $ r $ cannot describe the whole spacetime, as it contains coordinate singularity at the throat. Therefore, it is only valid to describe one side of the throat. So, we shall denote the whole spacetime by introducing proper distance $ l(r) $, such that the Eq. \eqref{math01} takes the form,
\begin{equation}\label{geo04}
	ds^2=- e^{2 \Phi(r(l))} dt^2+ dl^2+r(l)^2 d\Omega^2,
\end{equation}
where,
\begin{equation}\label{geo05}
	l(r)= \pm \int_{b_0}^{r} \left( 1-\frac{b(r)}{r} \right)^{-1/2} dr.
\end{equation}
Here, the position of the throat lies at $ l=0 $, and the two spatial infinities $ l \rightarrow \pm \infty $ define the two sides of the wormhole throat, where the sign of proper radius is positive and negative respectively for upper and lower universes. Notice that, if the wormhole describes a bridge between two different universes, then the upper universe can be considered as our universe where the lower one denotes other universe. Otherwise, if the geometry connects two distant parts of our universe, the upper and lower universe define our locality and the distant sublocality in the same universe. 

To find out the geodesic equations on the equatorial plane where $ \theta=\frac{\pi}{2} $, we shall consider the Lagrangian equation of motion of the wormhole metric defined in \eqref{math01} as, $ \mathcal{L} = \frac{1}{2} g_{\alpha \beta} \frac{dx^\alpha}{d\tau} \frac{dx^\beta}{d\tau} $. Hence, the first order geodesics are obtained from the canonical conserved momentum and are given by
\begin{eqnarray}
	\dot{t} &=& E e^{-2 \Phi(r)},
	\label{geo06} \\
	\dot{\phi} &=& L/r^2,
	\label{geo07} \\
	\dot{r}^2 &=& \left( 1-\frac{b(r)}{r} \right)\left(E^2 e^{-2 \Phi(r)} -\frac{L^2}{r^2}+\varepsilon \right),
	\label{geo08}
\end{eqnarray}
where $ \varepsilon=1, 0, -1 $ denotes the spacelike, null and timelike geodesics, and $ E, L $ are the energy and angular momentum of the particle, respectively. The overdot is the differentiation with respect to the affine parameter for null geodesics, and proper time for timelike geodesics.

For radial null geodesics, Eq. \eqref{geo08} becomes
\begin{eqnarray}
	\dot{r}^2 = \left( 1-\frac{b(r)}{r} \right)\left(E^2 e^{-2 \Phi(r)} -\frac{L^2}{r^2} \right).
	\label{geo09}
\end{eqnarray}

Here, to discuss the photon orbits, one may introduce,
\begin{eqnarray}
	\left( \frac{dr}{d\phi} \right)^2 = \left( 1-\frac{b(r)}{r} \right)\left( \frac{E^2}{L^2}r^4 e^{-2\Phi(r)}-{r^2} \right).
	\label{geo10}
\end{eqnarray}

Considering a source of photon radius $ r_s $ is causing the geometry, photons coming from spatial infinity never hit the surface if an existing solution obeys the condition $ r_0>r_s $, such that $ \dot{r}^2=0 $. We describe $ r_0 $ to be the turning point or the distance of closest approach \cite{mishra2018} for which,
\begin{eqnarray}
	\frac{L^2}{E^2}=r_0^2 e^{-2\Phi(r_0)}~~~~~~(\text{if}~\frac{b(r)}{r}\ne 1~\text{for~any}~r>r_s).
	\label{geo11}
\end{eqnarray}

Thus, the impact parameter takes the form
\begin{eqnarray}
	\mu=\frac{L}{E}=\pm r_0 e^{-\Phi(r_0)}.
	\label{geo12}
\end{eqnarray}

If a photon coming from spatial infinity, i.e. $ \lim\limits_{r \rightarrow \infty} \left( r, -\frac{\pi}{2}-\frac{\alpha}{2} \right) $, passes through the turning point $ (r_0,0) $ and reach  $ \lim\limits_{r \rightarrow \infty} \left( r, \frac{\pi}{2}+\frac{\alpha}{2} \right) $, then the deflection angle $ \alpha(r_0) $ becomes \cite{amrita2010},
\begin{equation}
	\alpha(r_0)=-\pi+2 \int_{r_0}^{\infty}
	\frac{dr}{r \left[ \left( 1-\frac{b(r)}{r} \right)\left( \left(\frac{r e^{\Phi(r_0)}}{r_0 e^{\Phi(r)}} \right)^2 -1 \right) \right]^{1/2}}.
	\label{geo13}
\end{equation}

\begin{figure}[t!]
	\centerline{\includegraphics[scale=.55]{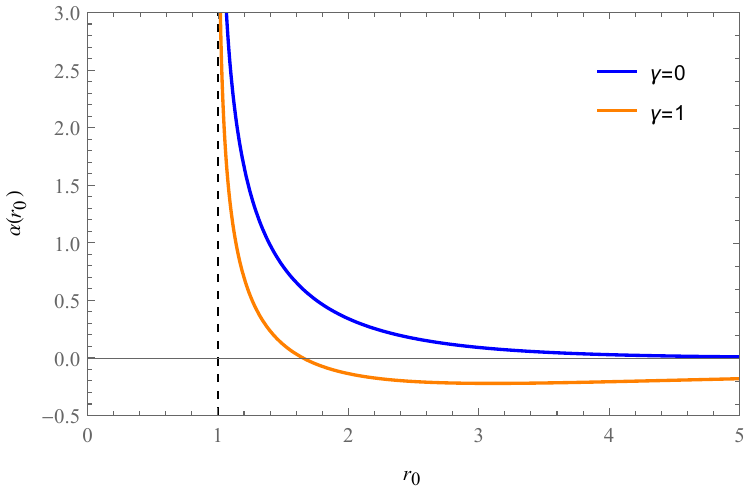}}
	\caption{Variation of the numerically integrated deflection angle against $ r_0 $. The plots transit to negative deflection angle at $ r_0=4.05,~\text{and}~r_0=1.65 $ respectively for $ \gamma=0,~\text{and}~\gamma=1 $. The vertical dashed line represents the throat radius at $ b_0=1 $. }
	\label{deflec_plot}
\end{figure}

The photon sphere exists for which the photons get trapped inside a sphere having fixed $ r $, and could not approach $ \lim\limits_{r \rightarrow \infty} \left( r, \frac{\pi}{2}+\frac{\alpha}{2} \right)
$, and the integral diverges. Introducing the shape and redshift function from Eq. \eqref{shape-eq} and \eqref{red-eq}, the deflection angle of Eq. \eqref{geo13} has been integrated numerically. A plot of the variation of deflection angle with $ r_0 $ for throat radius $ b_0=1 $ is shown in Fig. \ref{deflec_plot}. Notice that the deflection angle attains negative value after a certain distance, both for the zero and non-zero tidal force wormholes. This phenomenon represents the presence of repulsive effect of gravity on the geometry. It must be noted that repulsive gravity nature is quite well-known in modified gravity theories with exotic matters and phantom energies \cite{nakashi2019,kitamura2013,izumi2013,kitamura2014,nakajima2014,shaikh2017}. Recently, it is observed that the dRGT massive gravity \cite{derham2010,derham2011} is a common source of the effect where massive gravitons may play a vital role in the phenomenon \cite{dutta2023}. In the form of negative deflection angle, the repulsive effect of gravity was investigated by Panpanich \textit{et al.} \cite{panpanich2019}. However, the effect is not caused by the introduction of exponential shape function. Rather, for the choice of redshift function (in Eq. \eqref{red-eq}), one may readily verify the negative deflection angle with well-known shape functions, such as the Ellis wormhole shape. Comparing the values of transition points for $ \gamma=0,~\text{and}~\gamma=1 $, one can conclude that the repulsive effect is higher for wormhole geometries with a finite tidal force.

\subsection{Timelike geodesics}\label{time-geodesic}
In this subsection, first we shall discuss a few key motions of massive particles around wormhole geometry. For ultrastatic wormholes, initial velocity, position and accelerations of radial timelike geodesics were briefly discussed by Cataldo \textit{et al.} \cite{cataldo2017}. Now, the complete (non-radial) timelike geodesic accelerations can be obtained by differentiating the first order radial timelike geodesic equation given by Eq. \eqref{geo08}.
\begin{equation}
	\ddot{r}= \left(1-e^{-(r-b_0)} \right) \frac{L^2}{r^3}+ \frac{1}{2} \left(E^2-\frac{L^2}{r^2}-1 \right) e^{-(r-b_0)} .
	\label{geo14}
\end{equation}

At this point, we are about to study the circular timelike geodesics which has some particular point of interest. For usual cases, it is well established that circular timelike geodesics can only exist at the throat and for them radial velocity vanishes, that means $ \dot{r}=0 $. Here, the acceleration for $ \dot{r}=0 $ is
\begin{equation}
	\ddot{r}= \left(1-e^{-(r-b_0)} \right) \frac{L^2}{r^3} .
	\label{geo15}
\end{equation}
As at $ r=b_0 $, the acceleration vanishes; one can always verify that circular timelike geodesics can only exist at the throat.

For ultrastatic wormholes, the radial geodesic equation in terms of proper distance can be written as
\begin{equation}
	\left( \frac{dl}{ds} \right)^2 = E^2-V(L,l) ,
	\label{geo16}
\end{equation}
where the potential related to conserved momentum and proper distance is given by
\begin{equation}
	V(L,l)= \frac{L^2}{r(l)^2}+\varepsilon ,
	\label{geo17}
\end{equation}
which is always positive for $ L \ne 0 $ and goes to zero for $ l \rightarrow \pm \infty $. Here,
\begin{equation}
	\frac{dV}{dl}\Big|_{l=0}=0 ,
	\label{geo18}
\end{equation}
and
\begin{equation}
	\frac{d^{2}V}{dl^{2}}\Big|_{l=0}=-\frac{L^{2}}{r^{4}} \left( \frac{b(r)}{r}-b'(r) \right)\Big|_{l=0}<0 .
	\label{geo19}
\end{equation}

Eq. \eqref{geo18} describes that the potential possesses a global maximum at the throat, and the inequality in Eq. \eqref{geo19} results from the flaring out condition. If $ r(l) $ concave up, the only turning point is given by $ l=0 $ which differentiates the bound orbits from the transmitting unbounded orbits. We have some stable circular geodesic orbits if $ r(l) $ is not necessarily concave up, and the potential well has oscillatory motion around turning points \cite{taylor2014, sarbach2012}.

The timelike geodesic trajectories can always be classified into bounded and unbounded orbits. Whereas, geodesics coming from spatial infinity reach the throat and reflect back to the same spatial infinity are called as bound orbits; in unbounded trajectory, geodesics pass through the throat to reach other universes and thus other spatial infinity. In traversable wormholes, it is of the utmost importance that the massive (timelike) particle trajectory started from one universe crosses the throat to travel to the other universe (or at a distant point of the same universe). This unbounded trajectory is also termed as the escape orbit.

Recalling Eq. \eqref{geo08} in terms of proper distance, we have
\begin{equation}
	\dot{l}^2 = \left(E^2 e^{-2\Phi(r)}-\frac{L^2}{r^2}-1 \right) .
	\label{geo20}
\end{equation}

If the particle coming from infinity gets deflected after approaching the throat at a closest distance $ b_0 $, then the bound orbit condition appears to be,
\begin{equation}
	\left(E^2 e^{-2\Phi(b_0)}-\frac{L^2}{b_0^2}-1 \right) \le 0,
	\label{geo21}
\end{equation}
which is simplified to,
\begin{equation}
	\frac{L}{\sqrt{E^2 e^{-2\Phi(b_0)}-1}} \ge b_0.
	\label{geo22}
\end{equation}
Otherwise the trajectory is an escape orbit. Readers are referred to \cite{mishra2018} for a detailed study of bound and escape orbit conditions for different choices of wormholes.

Imposing the values of our choices, the bound orbit condition given by Eq. \eqref{geo22} is modified to
\begin{eqnarray}
	\frac{L}{\sqrt{E^2 e^{-\gamma/b_0}-1}} \ge b_0 ,
	\label{orb_con-tidal}
\end{eqnarray}
and for zero-tidal force wormhole, it is written as
\begin{eqnarray}
	\frac{L}{\sqrt{E^2-1}} \ge b_0 .
	\label{orb_con}
\end{eqnarray}

\section{The Geodesic Trajectories and CTG}\label{orbits}

\begin{figure*}[]
	\centering
	\subfloat[Escape orbit for $ E=15.633, L=15.6, b_0=1 $\label{escape-a}]{{\includegraphics[height=5cm]{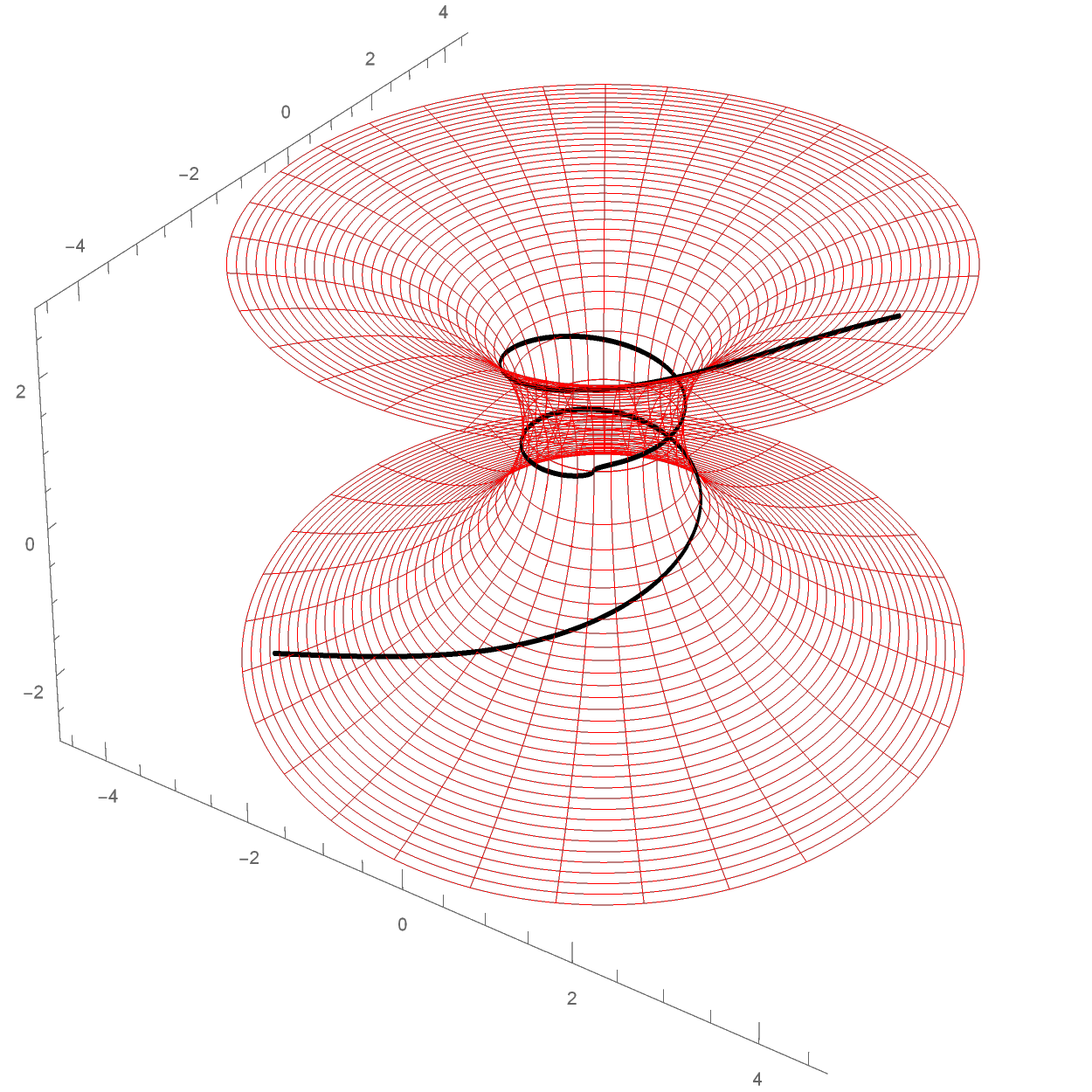}}}\qquad
	\subfloat[Escape orbit for $ E=15.64, L=15.6, b_0=1 $\label{escape-b}]{{\includegraphics[height=5cm]{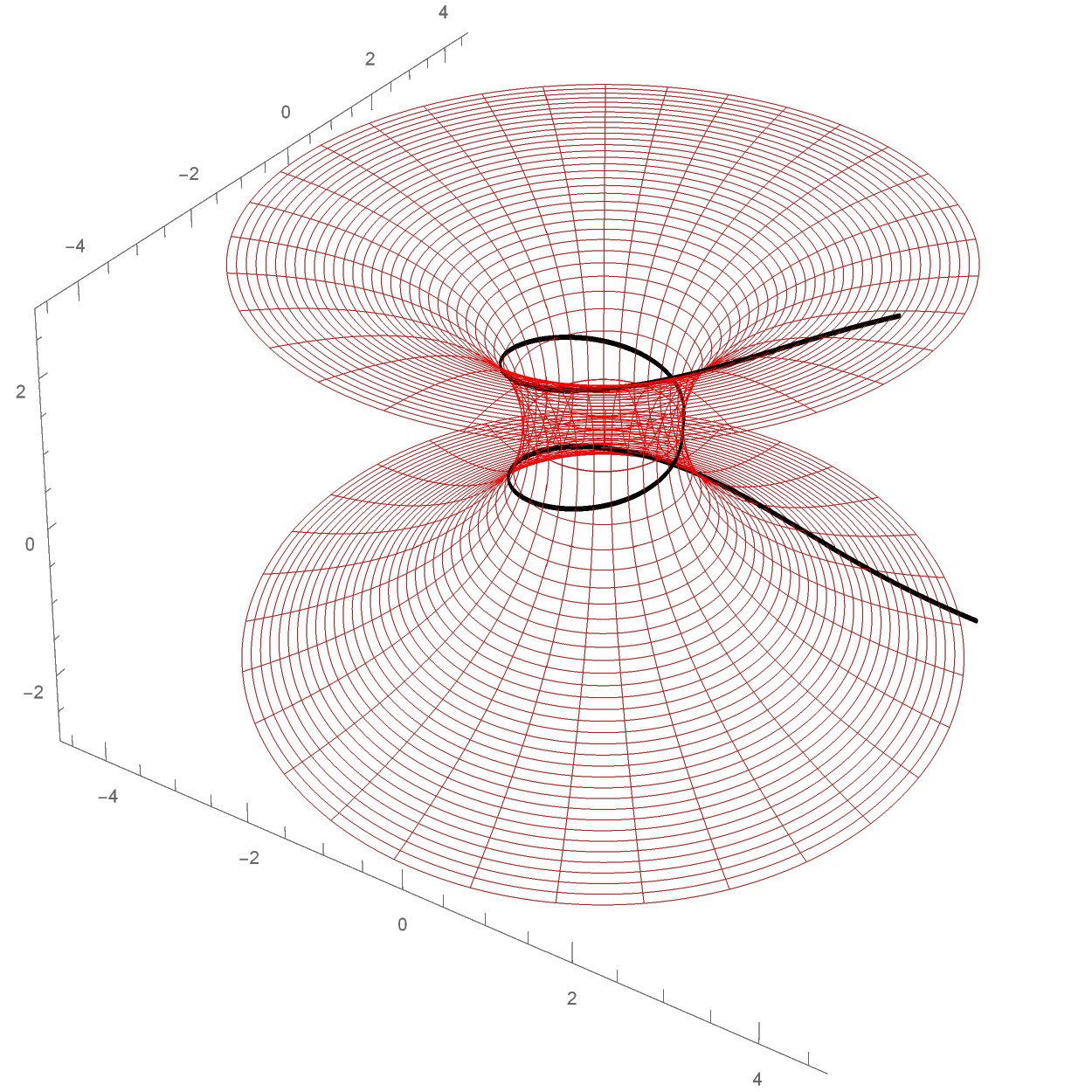}}}\qquad
	\subfloat[Escape orbit for $ E=15.7, L=15.6, b_0=1 $\label{escape-c}]{{\includegraphics[height=5cm]{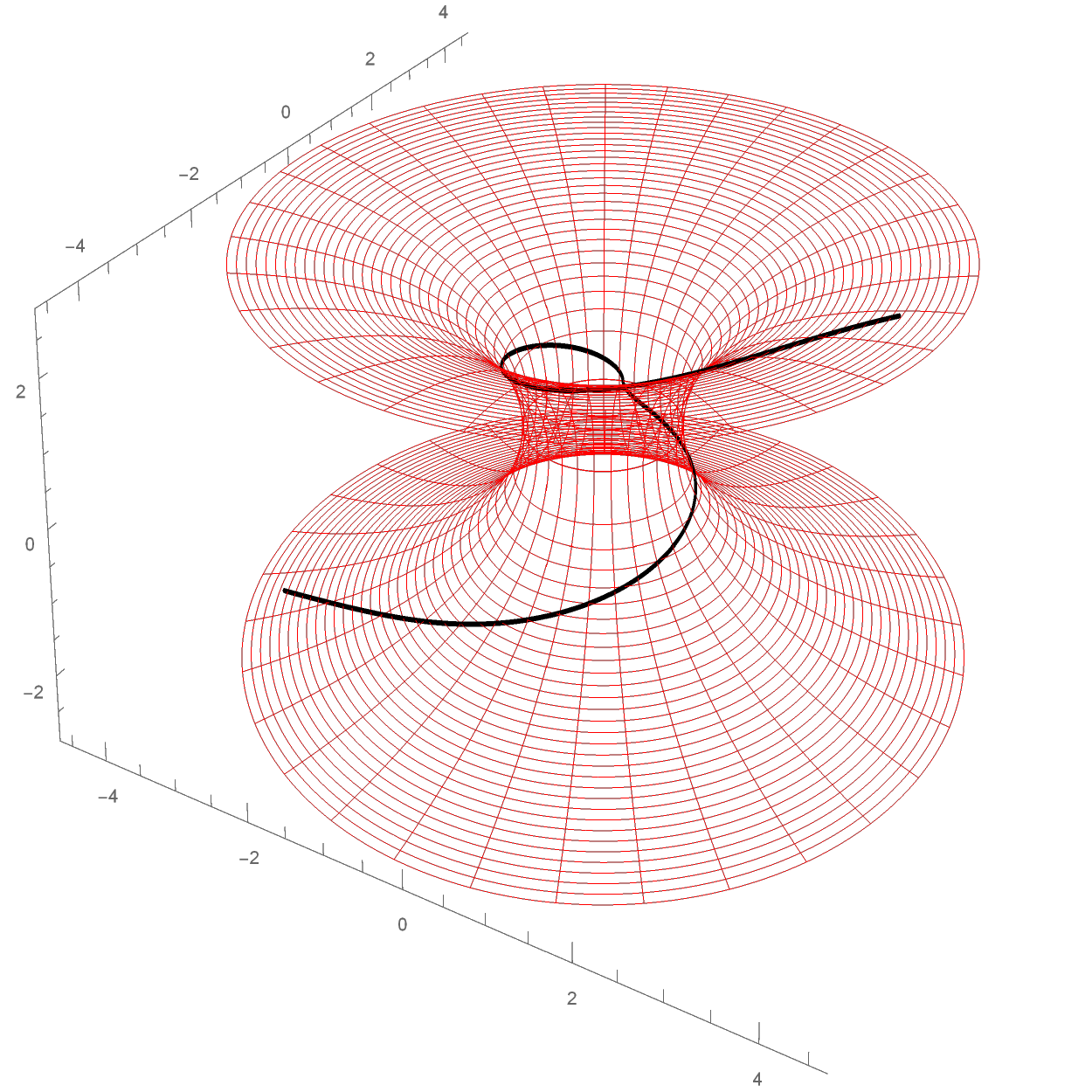}}}\\
	\subfloat[Escape orbit for $ E=16.1, L=15.6, b_0=1 $\label{escape-d}]{{\includegraphics[height=5cm]{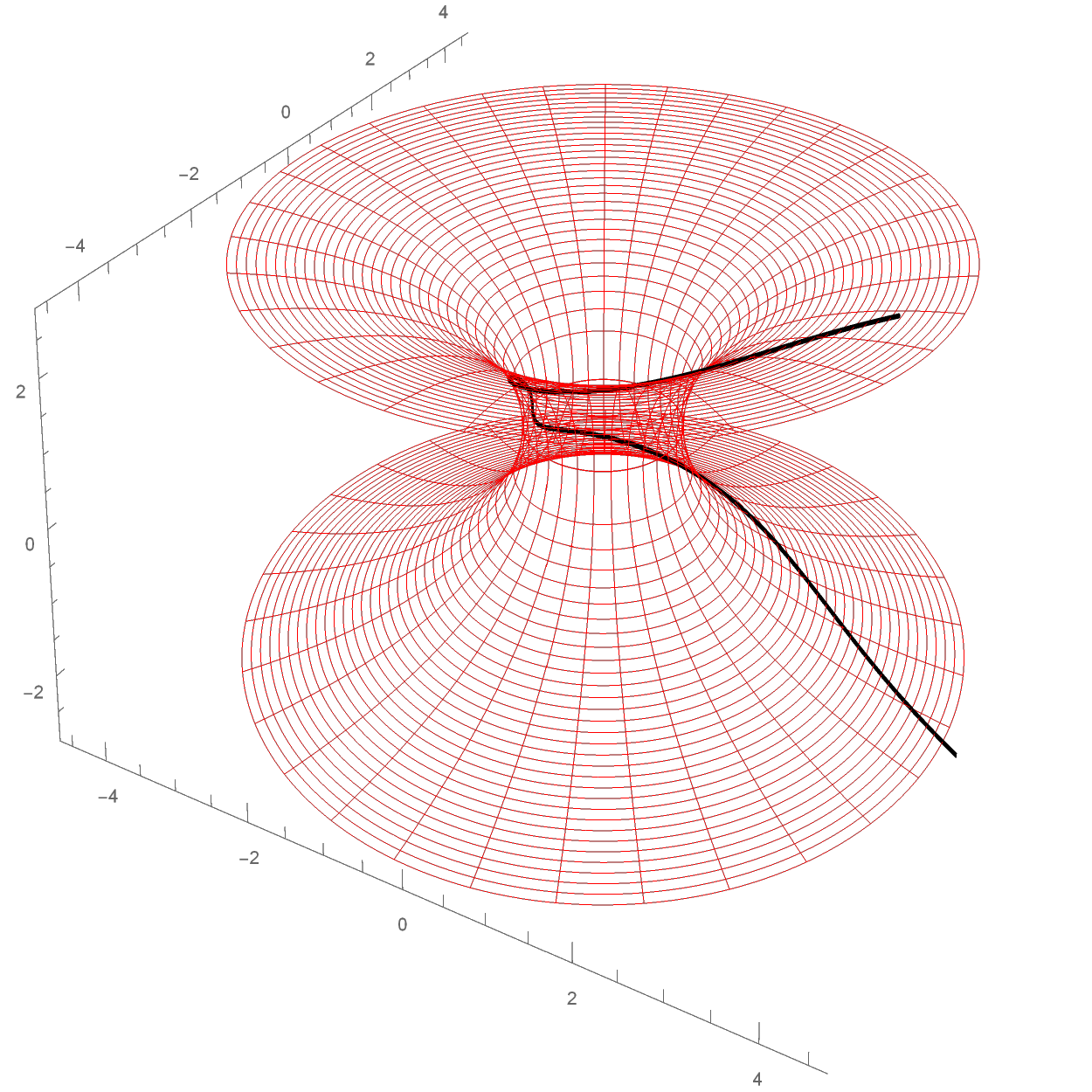}}}\qquad
	\subfloat[Escape orbit for $ E=20.5, L=15.6, b_0=1 $\label{escape-e}]{{\includegraphics[height=5cm]{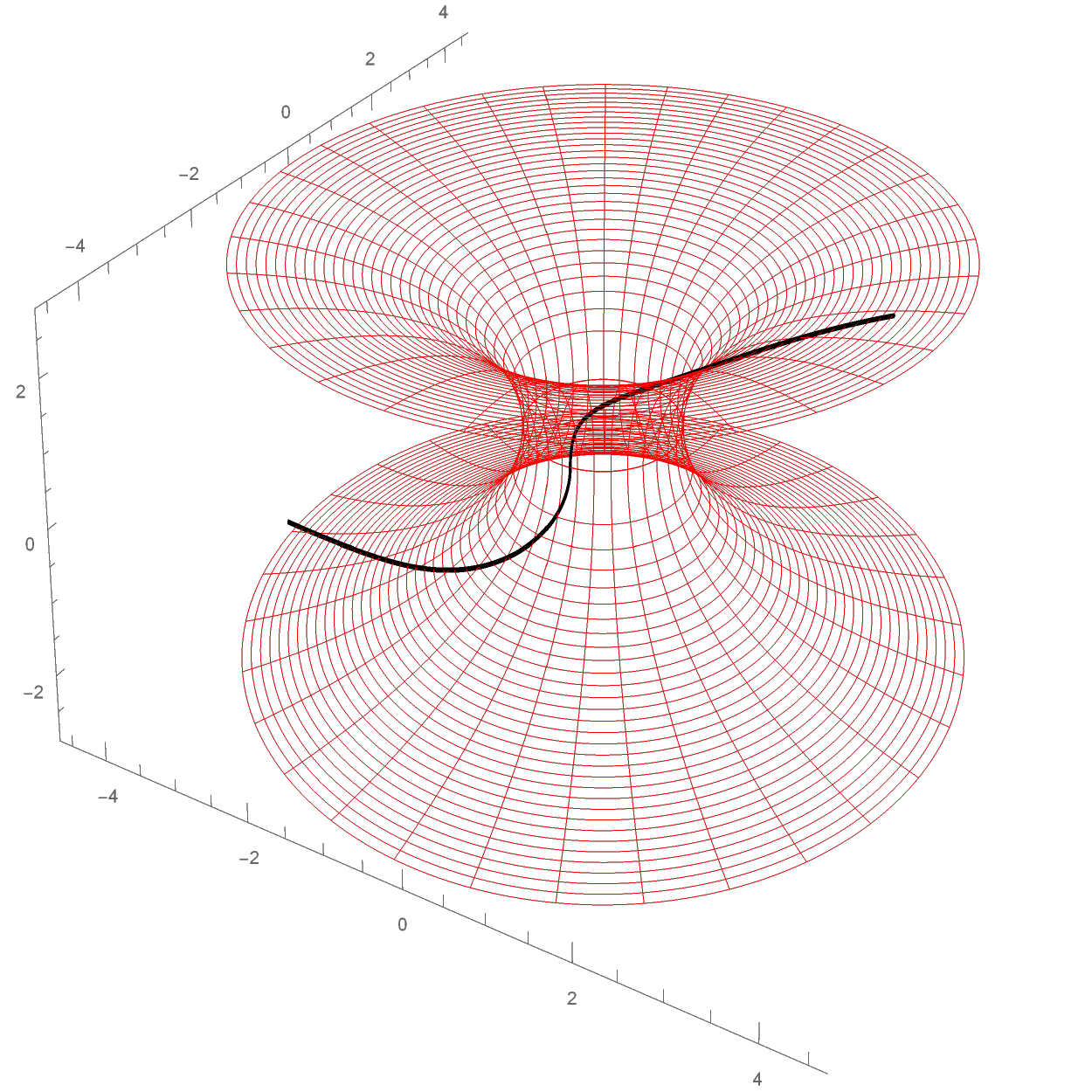}}}\qquad
	\subfloat[Escape orbit for $ E=23.5, L=15.6, b_0=1 $\label{escape-f}]{{\includegraphics[height=5cm]{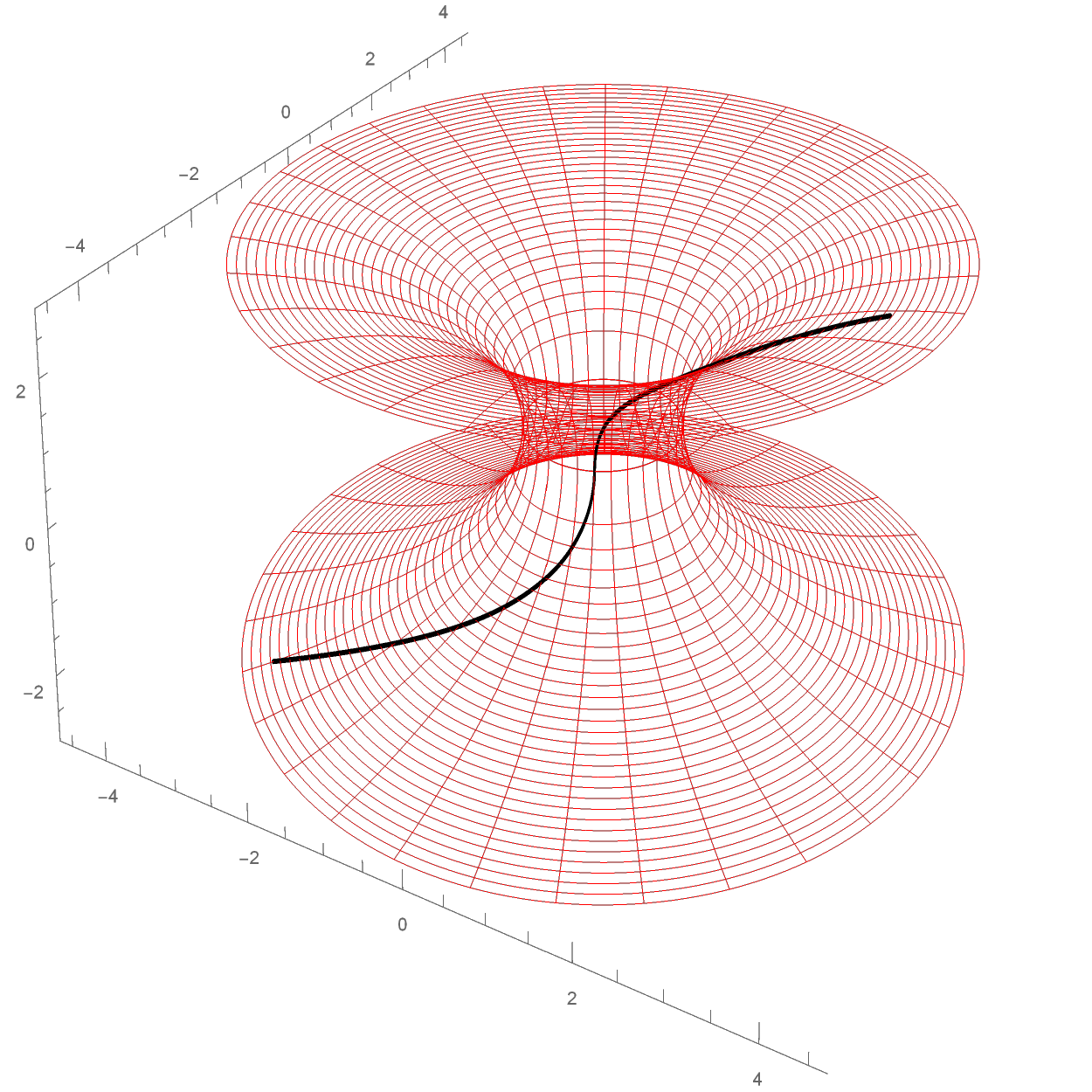}}}
	\caption{Escape orbits for zero-tidal force (i.e. $ \gamma=0 $) wormhole on the isometric embedding diagram with $ dl/d\phi $ solution for different values of parameters. The throat radius is chosen to be $ b_0=1 $. Variations of the trajectories are exhibited for gradually increasing energy ($ E $) where the angular momentum is fixed at $ L=15.6 $. Orbits start from the lower universe, cross the throat and pass to the upper universe.}
	\label{escape_plot}
\end{figure*}

\begin{figure*}[]
	\centering
	\subfloat[Bound orbit for $ E=10.5, L=15.6, b_0=1 $\label{bound-a}]{{\includegraphics[height=5cm]{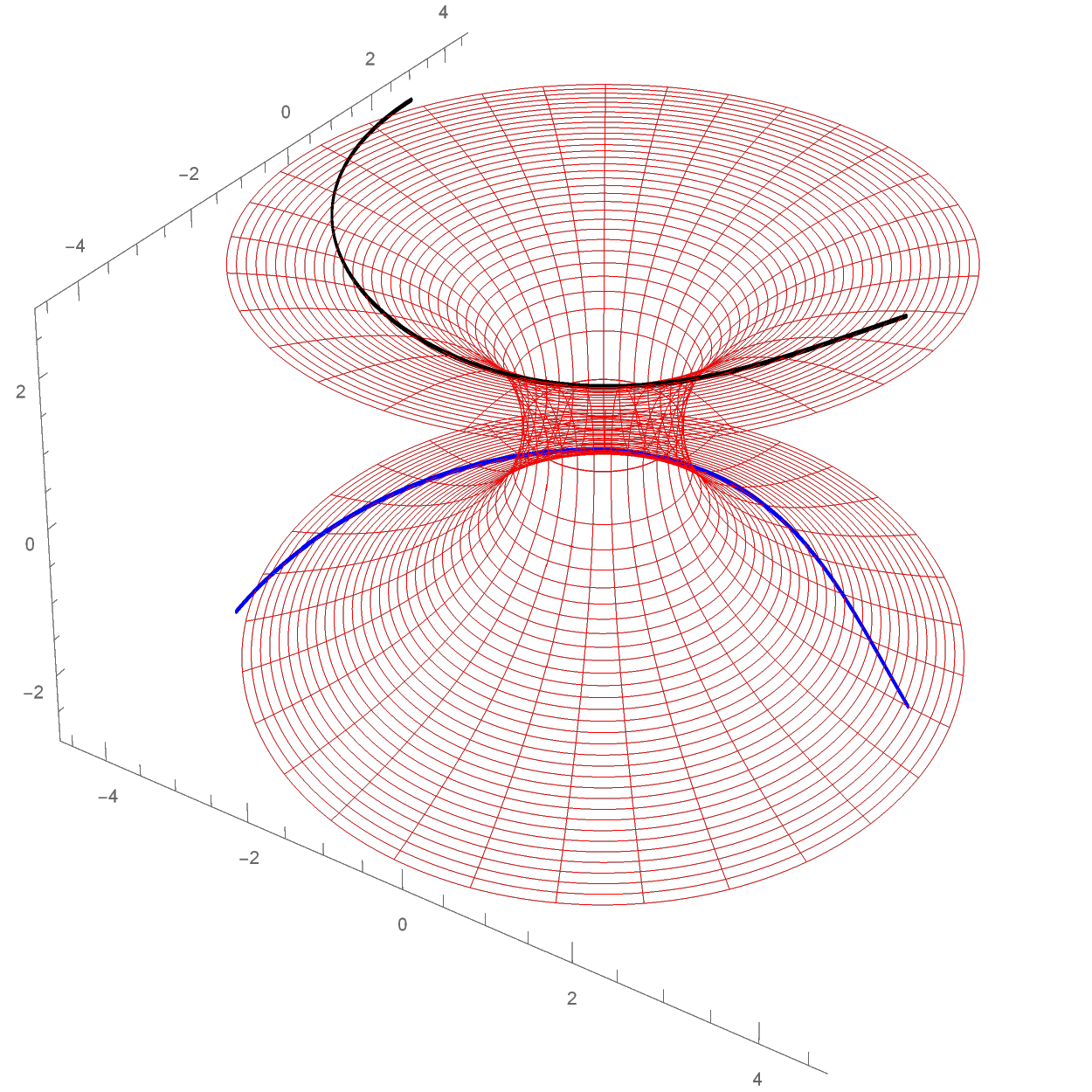}}}\qquad
	\subfloat[Bound orbit for $ E=12.5, L=15.6, b_0=1 $\label{bound-b}]{{\includegraphics[height=5cm]{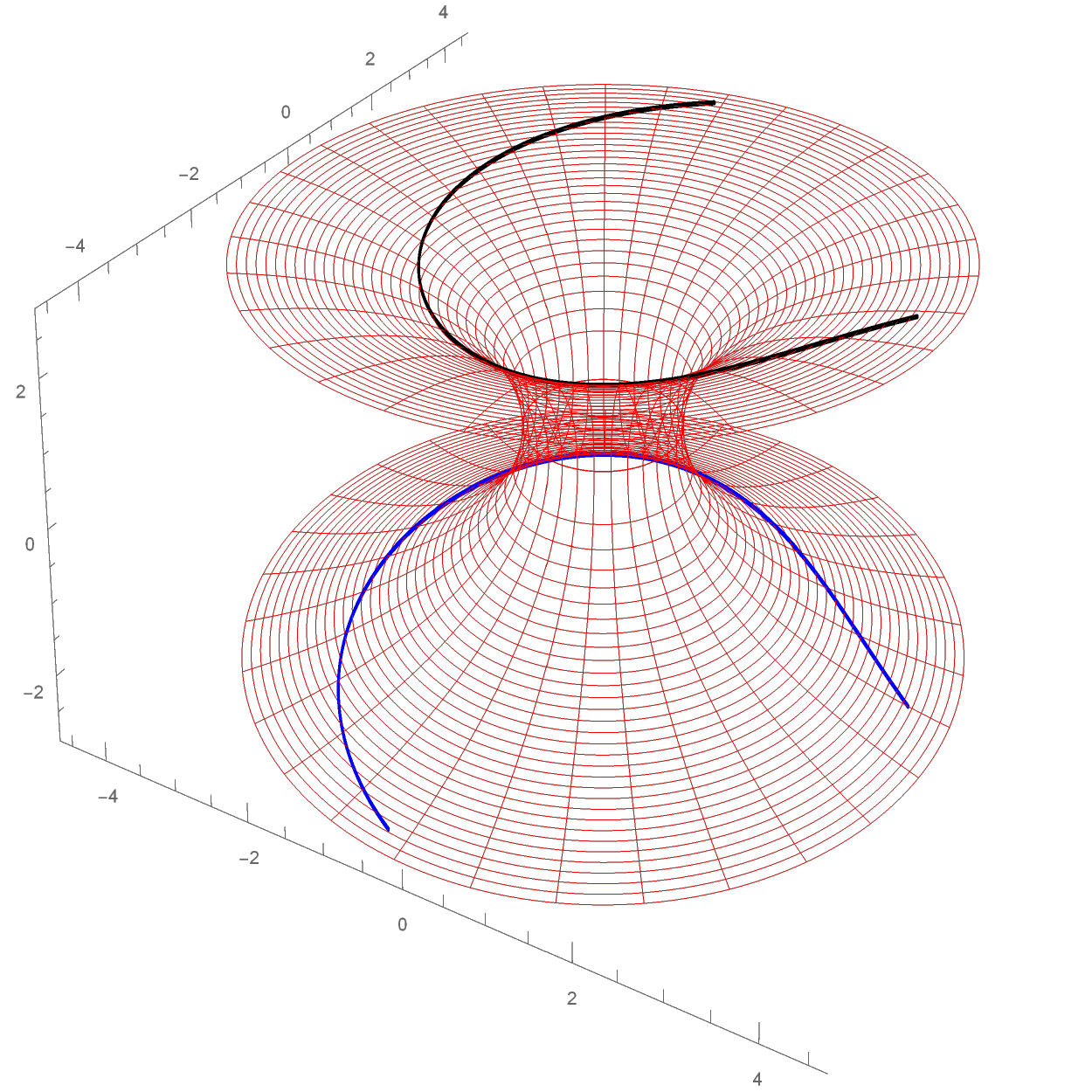}}}\qquad
	\subfloat[Bound orbit for $ E=14.5, L=15.6, b_0=1 $\label{bound-c}]{{\includegraphics[height=5cm]{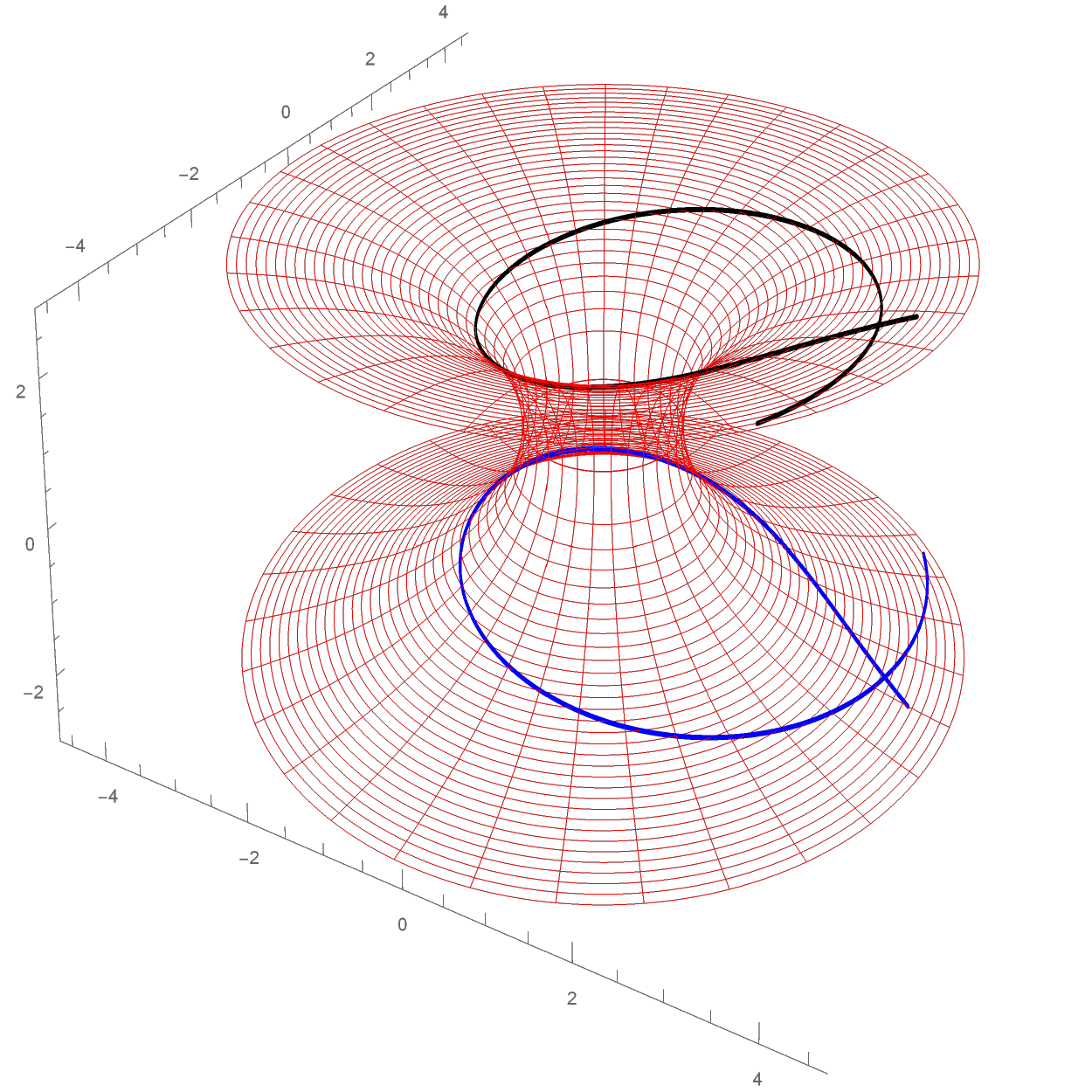}}}\\
	\subfloat[Bound orbit for $ E=15.5, L=15.6, b_0=1 $\label{bound-d}]{{\includegraphics[height=5cm]{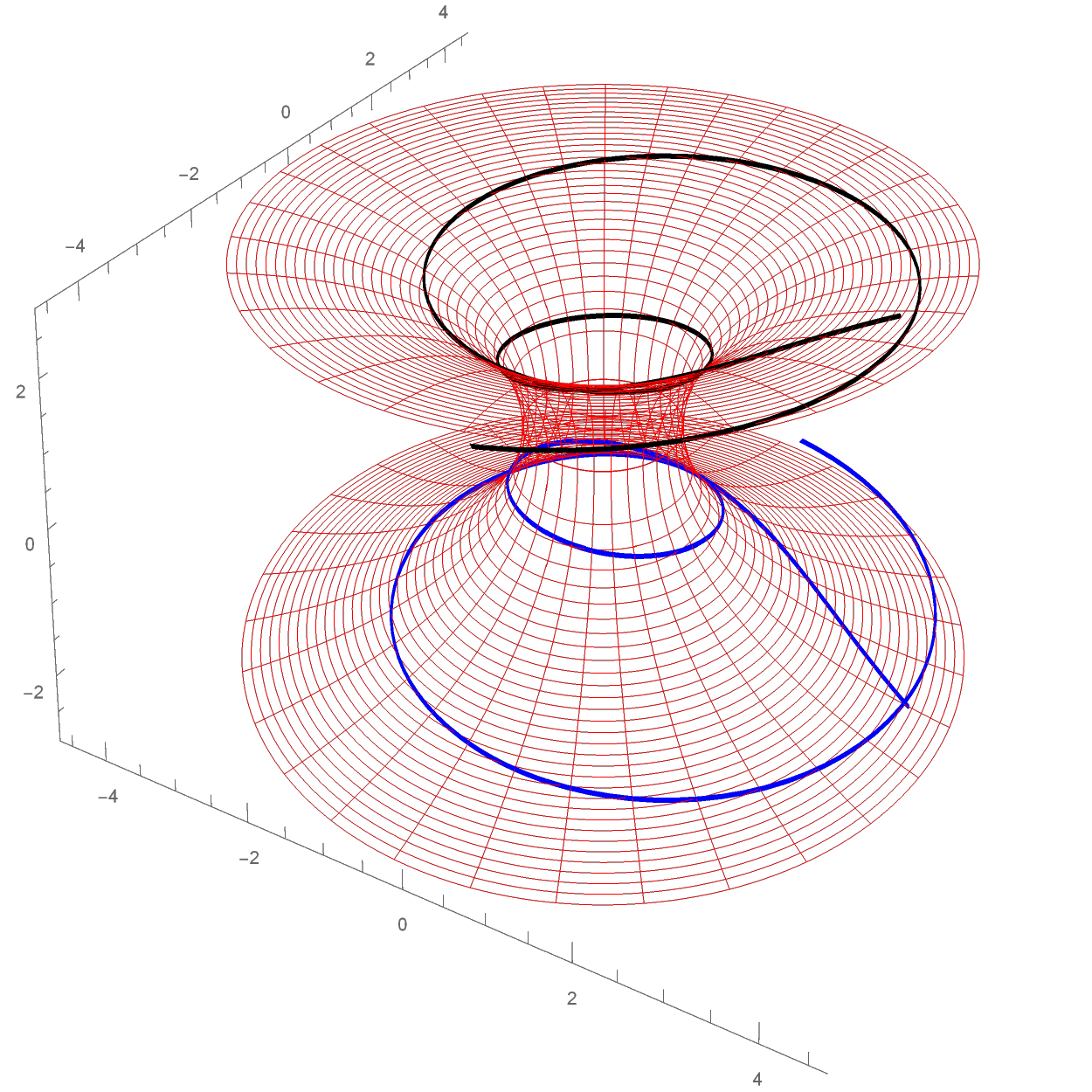}}}\qquad
	\subfloat[Bound orbit for $ E=15.6, L=15.6, b_0=1 $\label{bound-e}]{{\includegraphics[height=5cm]{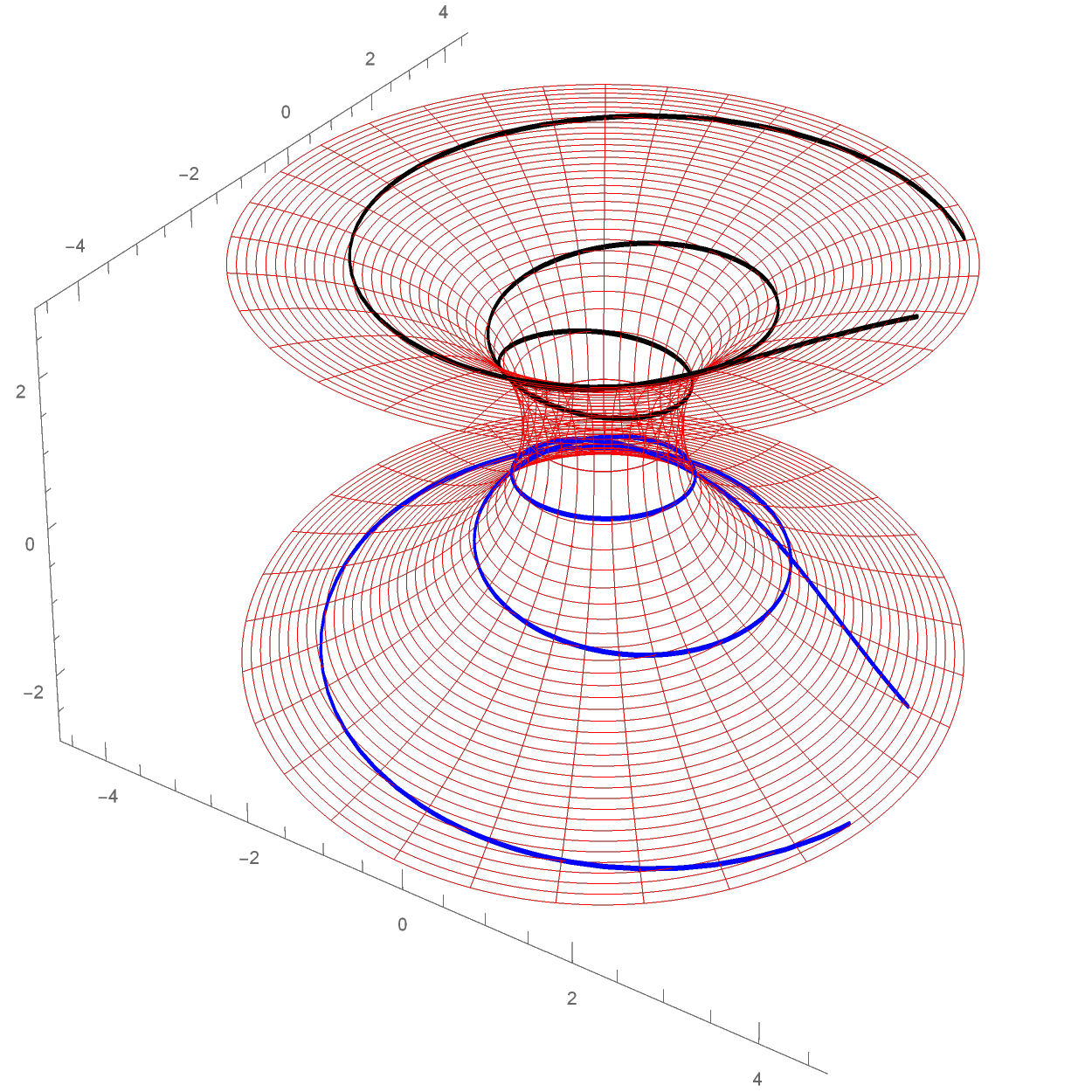}}}\qquad
	\subfloat[Bound orbit for $ E=15.63, L=15.6, b_0=1 $\label{bound-f}]{{\includegraphics[height=5cm]{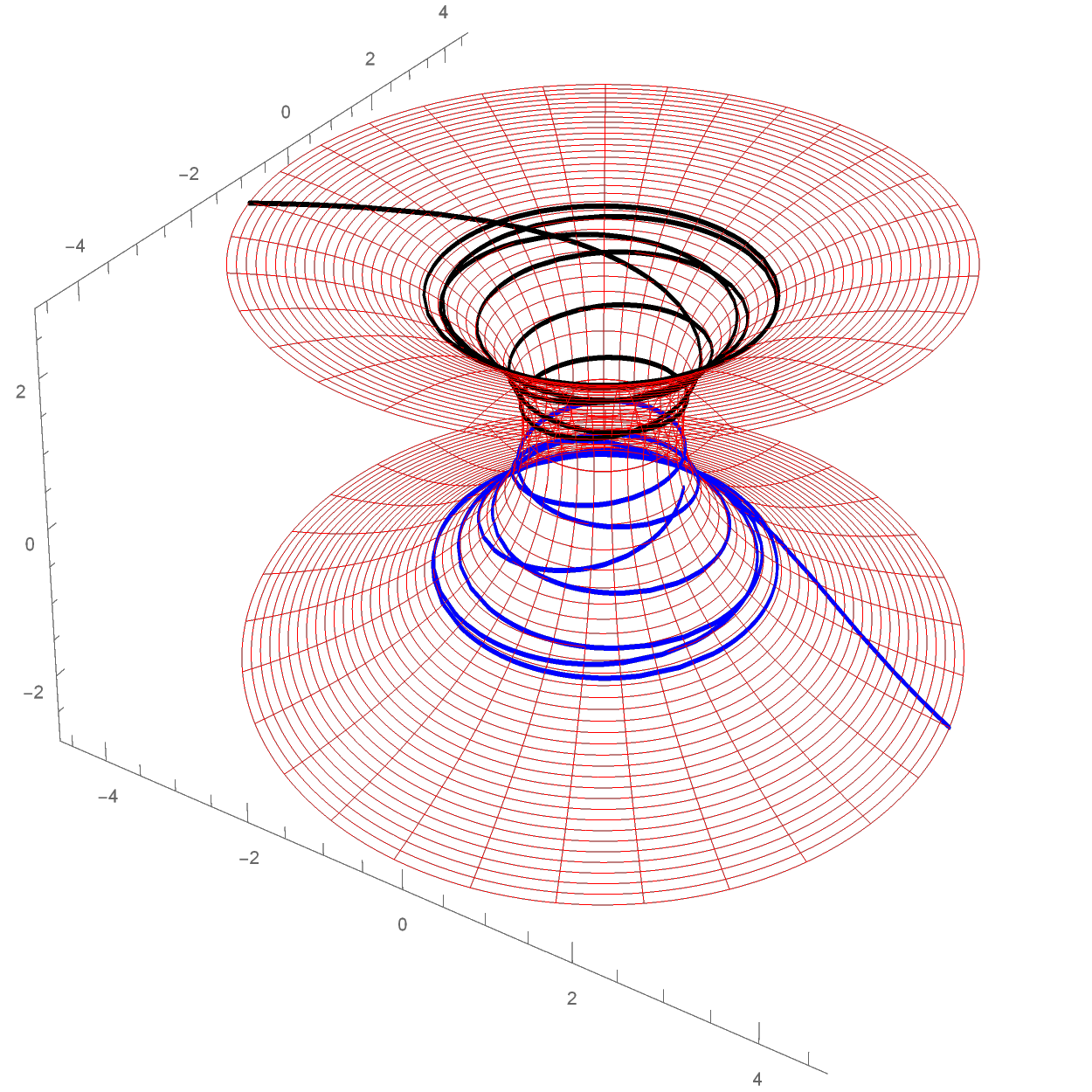}}}
	\caption{Bound orbits for zero-tidal force (i.e. $ \gamma=0 $) wormhole on the isometric embedding diagram with $ dl/d\phi $ solution for different values of parameters. The throat radius is chosen to be $ b_0=1 $. Variations of the trajectories are exhibited for gradually increasing energy ($ E $) where the angular momentum is fixed at $ L=15.6 $. Orbits started from the upper/lower universe, deflected from the throat region and returned to the same universe. The black and blue trajectories in the upper and lower universe respectively denote the results of positive and negative segments of $ dl/d\phi $. The orbits in \ref{bound-c}-\ref{bound-f} contain intersecting closed timelike geodesic trajectories.}
	\label{bound_plot}
\end{figure*}

\begin{figure*}[]
	\centering
	\subfloat[Escape orbit for $ E=25.773, L=15.6, b_0=1 $\label{tidal_escape-a}]{{\includegraphics[height=5cm]{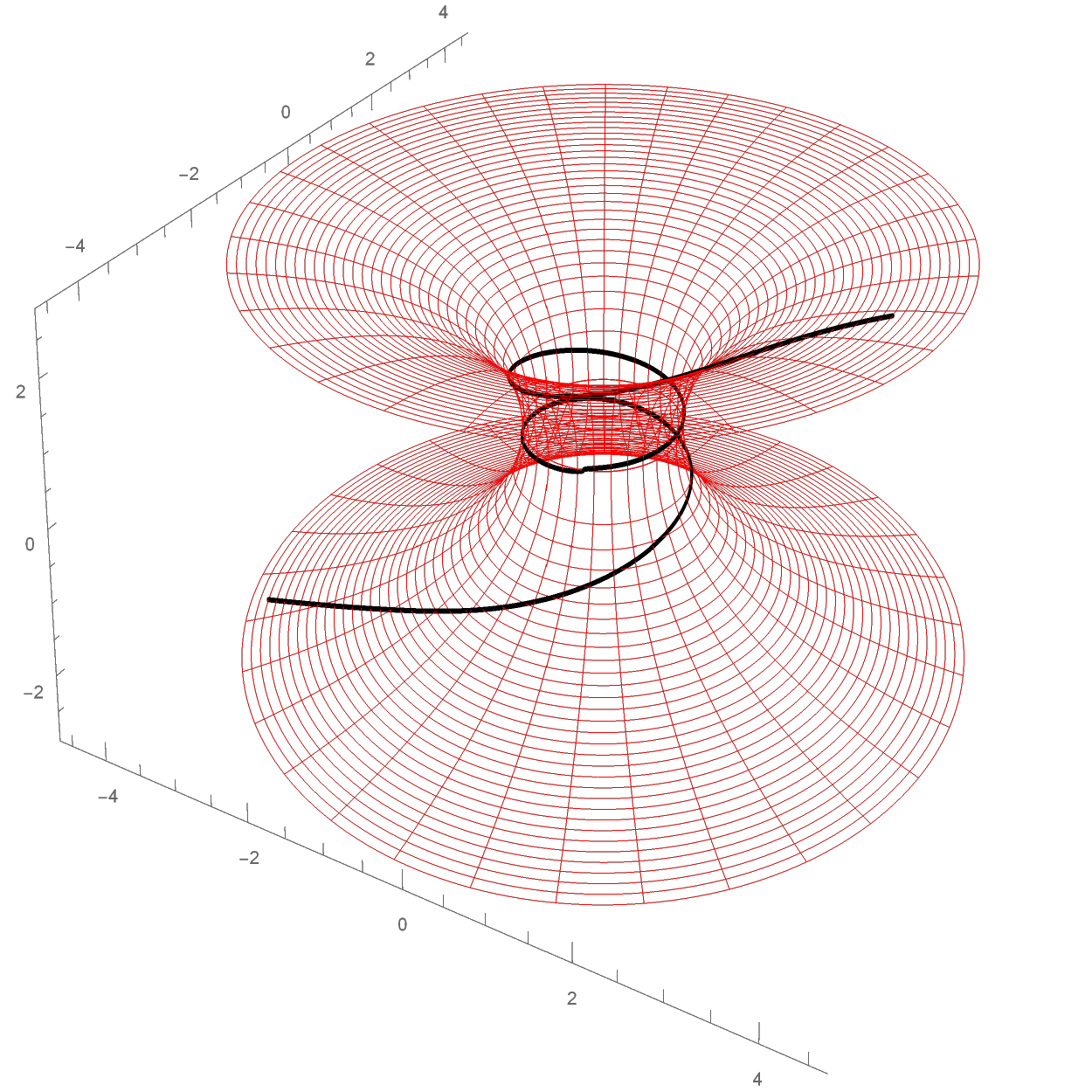}}}\qquad
	\subfloat[Escape orbit for $ E=25.78, L=15.6, b_0=1 $\label{tidal_escape-b}]{{\includegraphics[height=5cm]{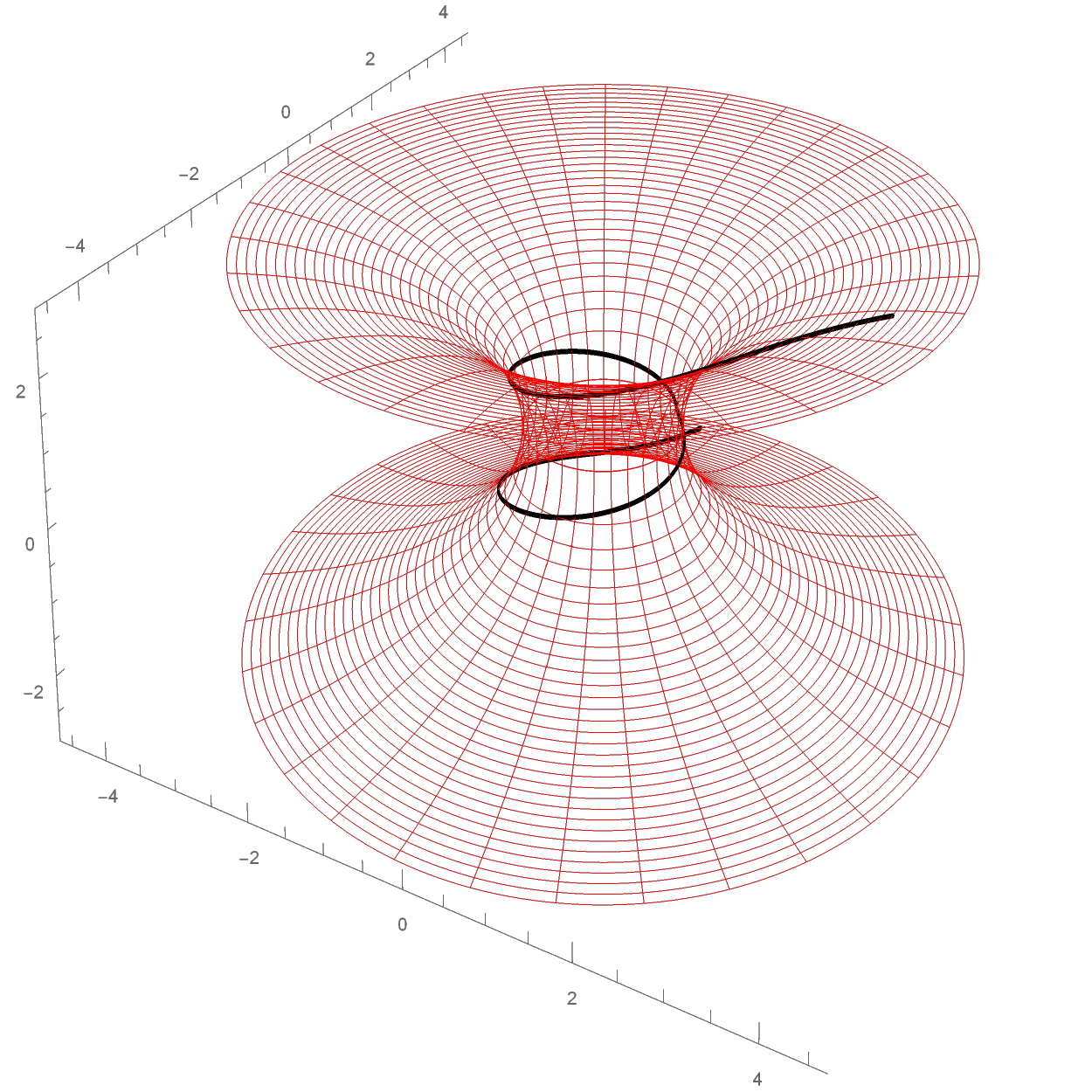}}}\qquad
	\subfloat[Escape orbit for $ E=25.9, L=15.6, b_0=1 $\label{tidal_escape-c}]{{\includegraphics[height=5cm]{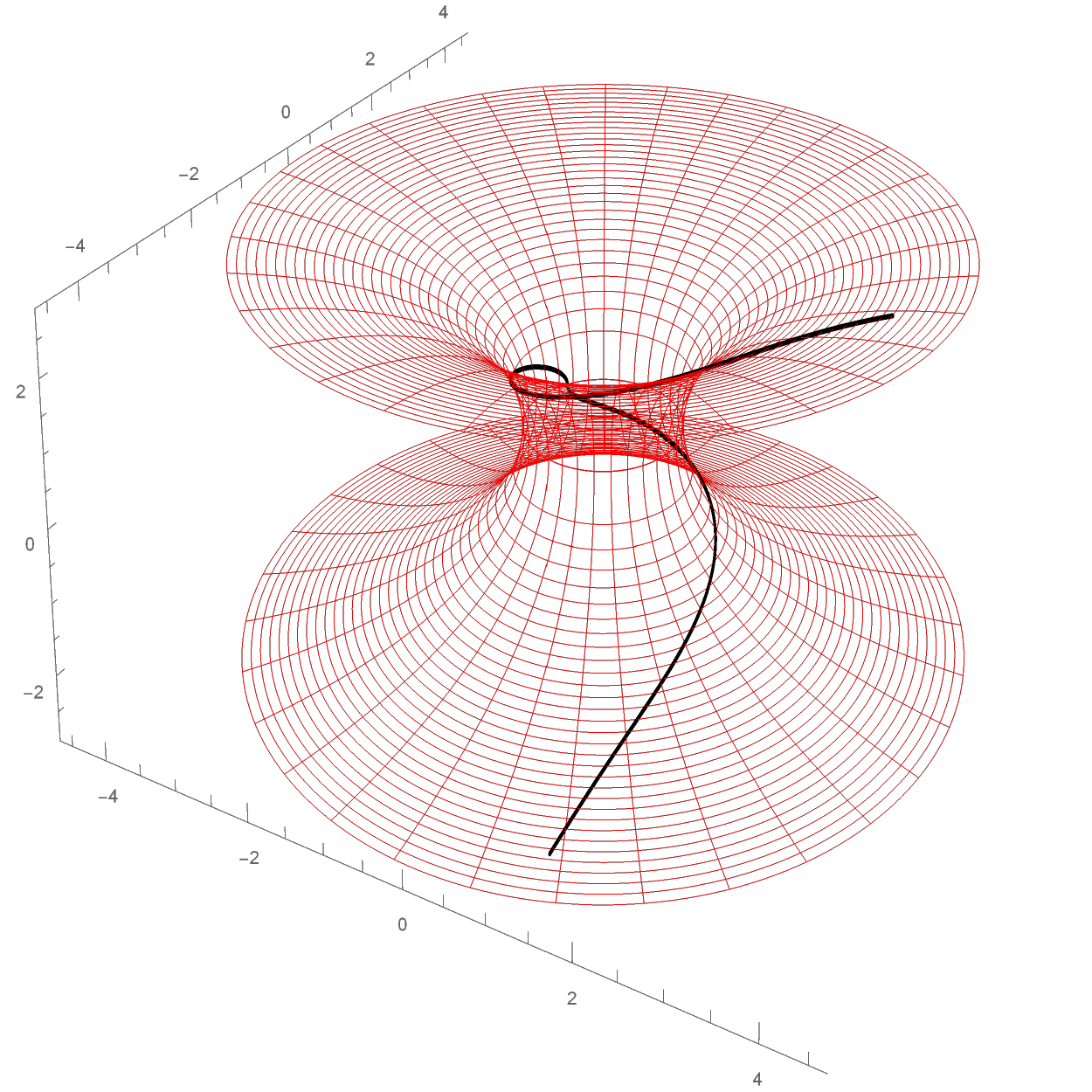}}}\\
	\subfloat[Escape orbit for $ E=26.5, L=15.6, b_0=1 $\label{tidal_escape-d}]{{\includegraphics[height=5cm]{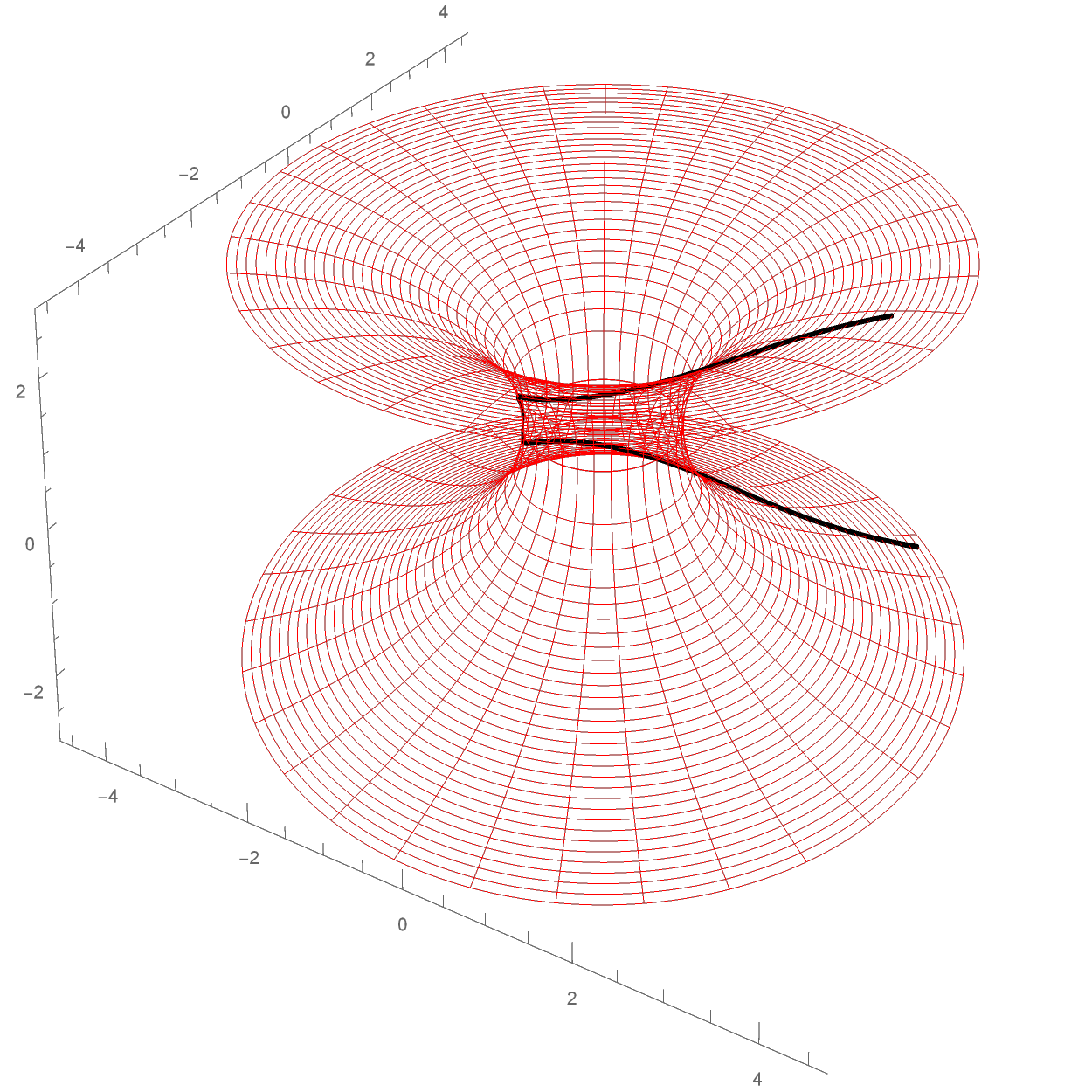}}}\qquad
	\subfloat[Escape orbit for $ E=30.5, L=15.6, b_0=1 $\label{tidal_escape-e}]{{\includegraphics[height=5cm]{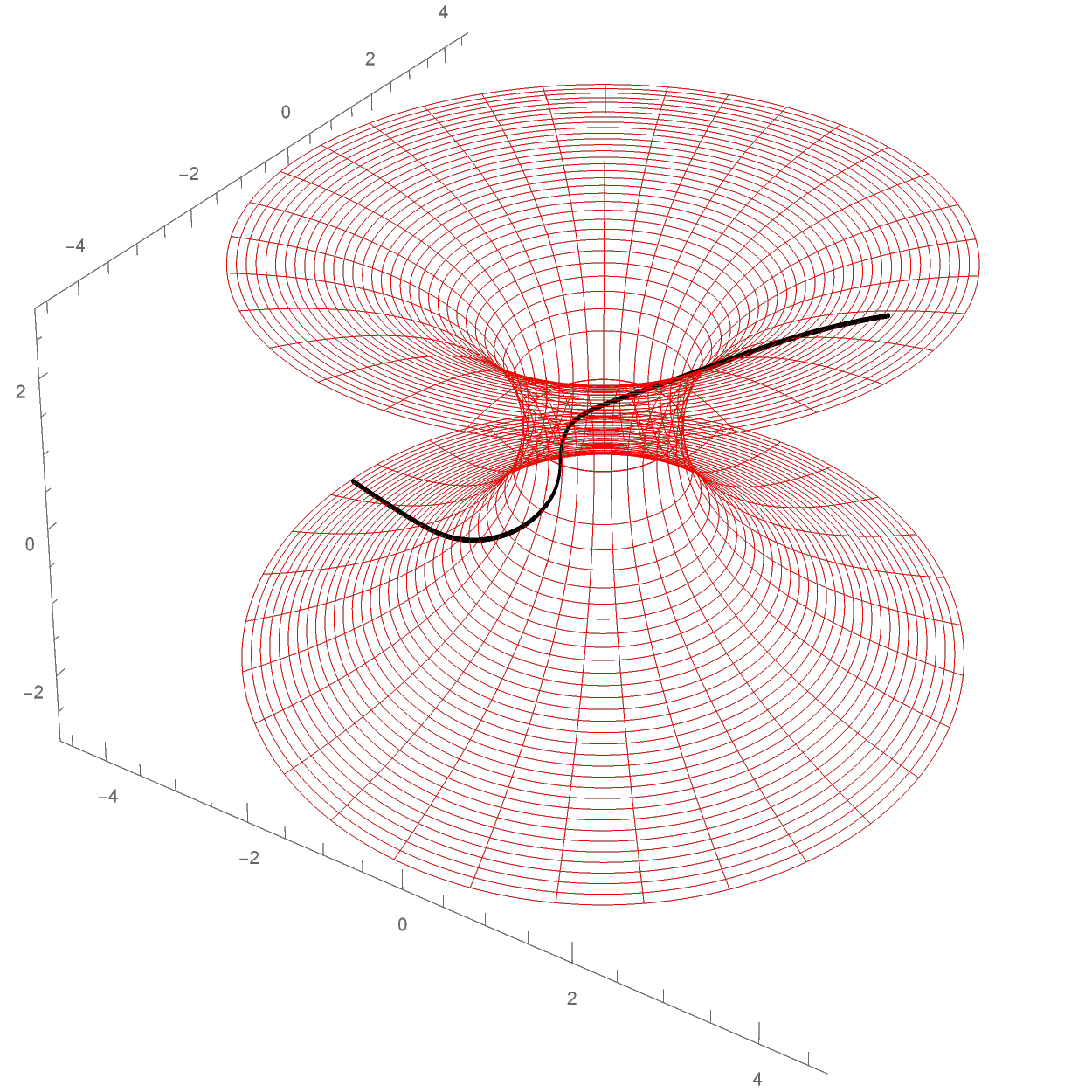}}}\qquad
	\subfloat[Escape orbit for $ E=35.5, L=15.6, b_0=1 $\label{tidal_escape-f}]{{\includegraphics[height=5cm]{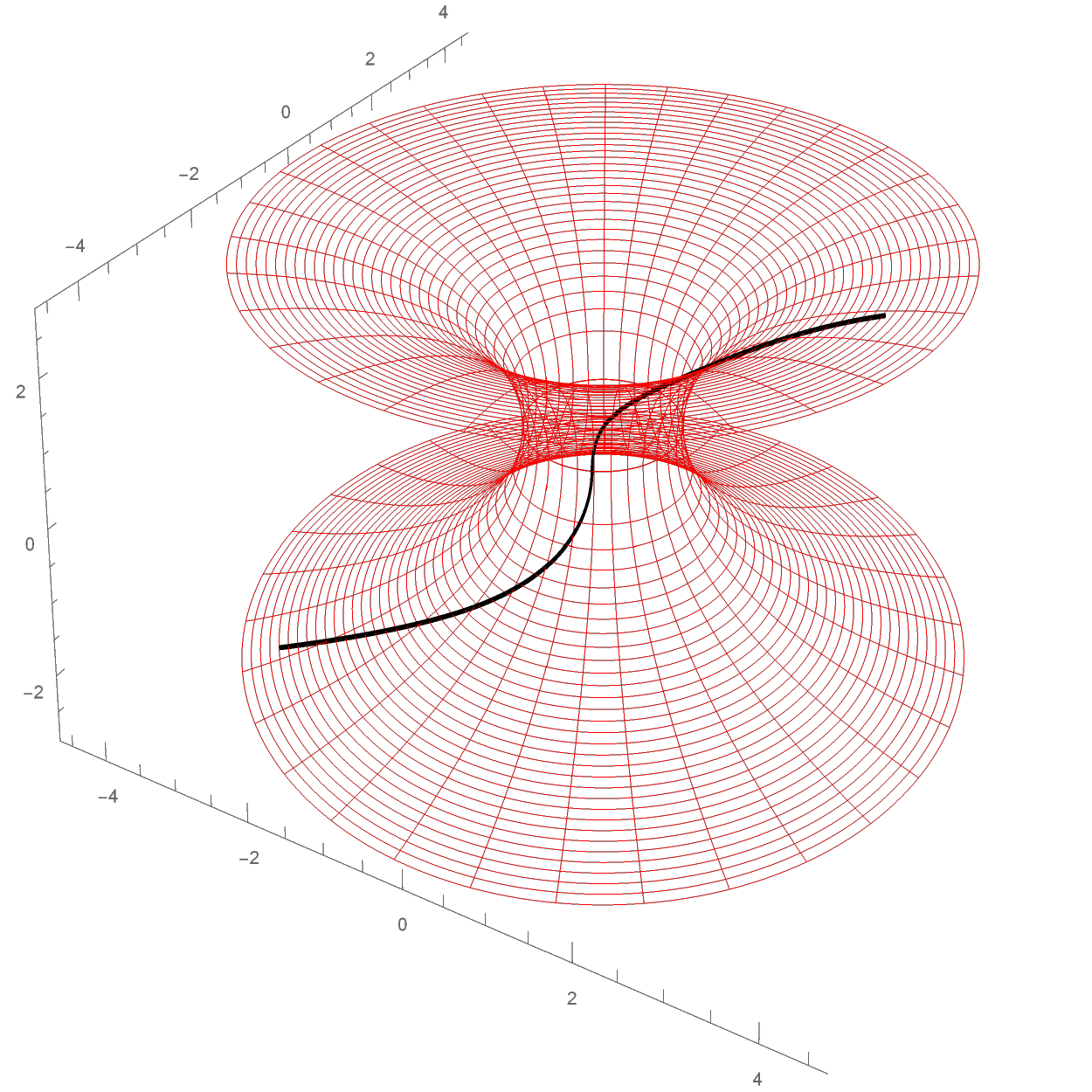}}}
	\caption{Escape orbits for non-zero tidal force (i.e. $ \gamma=1 $) wormhole on the isometric embedding diagram with $ dl/d\phi $ solution for different values of parameters. The throat radius is chosen to be $ b_0=1 $. Variations of the trajectories are exhibited for gradually increasing energy ($ E $) where the angular momentum is fixed at $ L=15.6 $. Orbits start from the lower universe, cross the throat and pass to the upper universe.}
	\label{tidal-escape_plot}
\end{figure*}

\begin{figure*}[]
	\centering
	\subfloat[Bound orbit for $ E=17.5, L=15.6, b_0=1 $\label{tidal_bound-a}]{{\includegraphics[height=5cm]{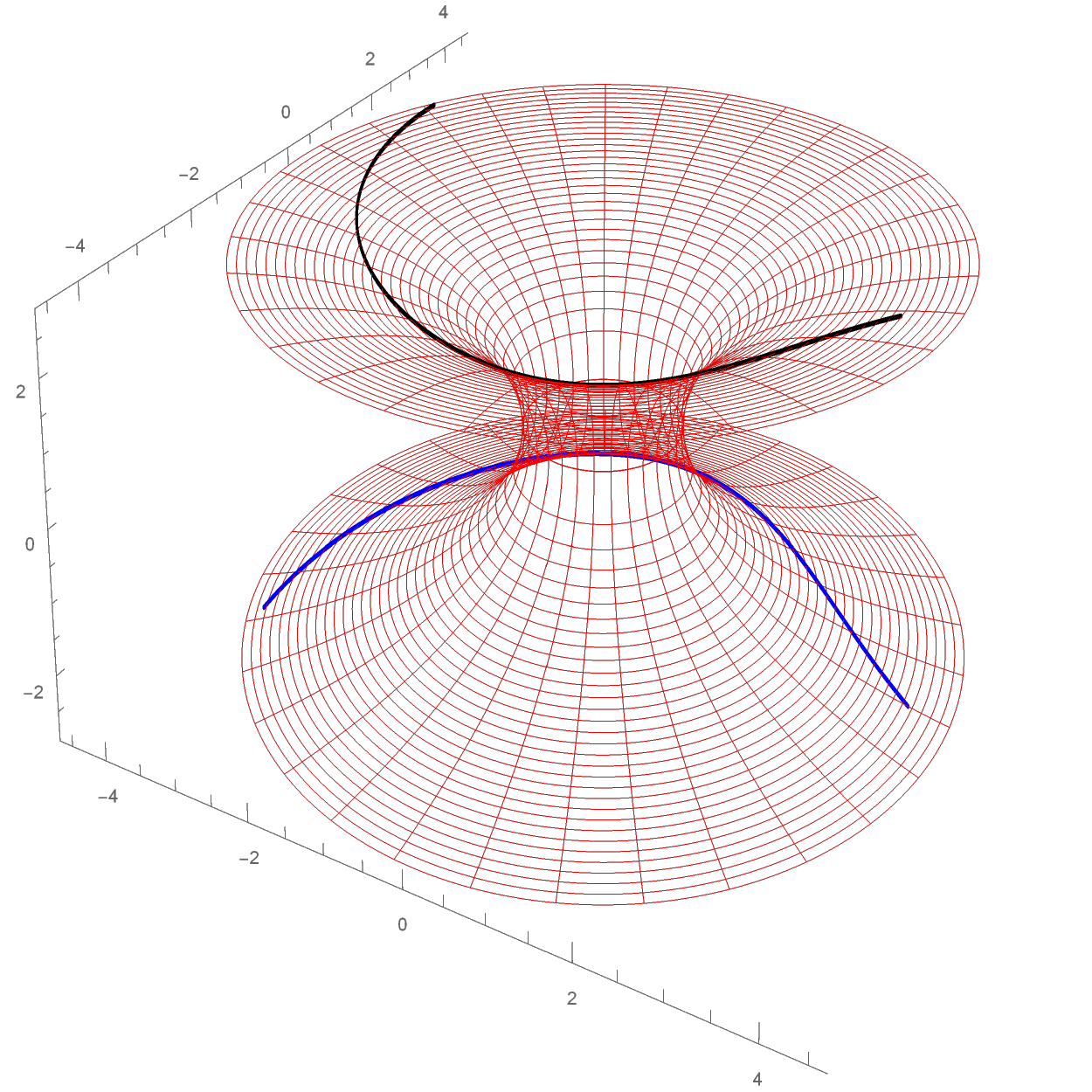}}}\qquad
	\subfloat[Bound orbit for $ E=20.5, L=15.6, b_0=1 $\label{tidal_bound-b}]{{\includegraphics[height=5cm]{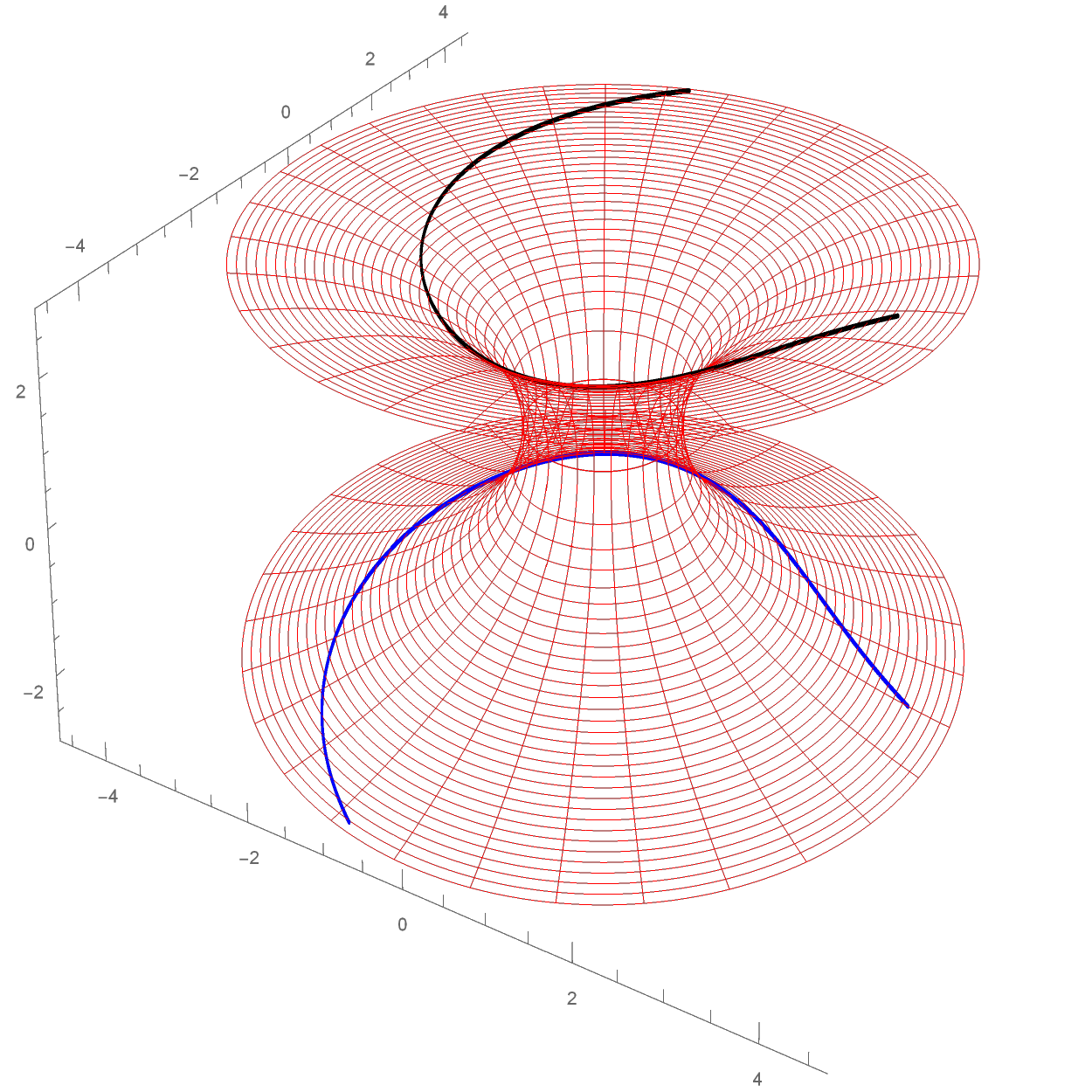}}}\qquad
	\subfloat[Bound orbit for $ E=25.5, L=15.6, b_0=1 $\label{tidal_bound-c}]{{\includegraphics[height=5cm]{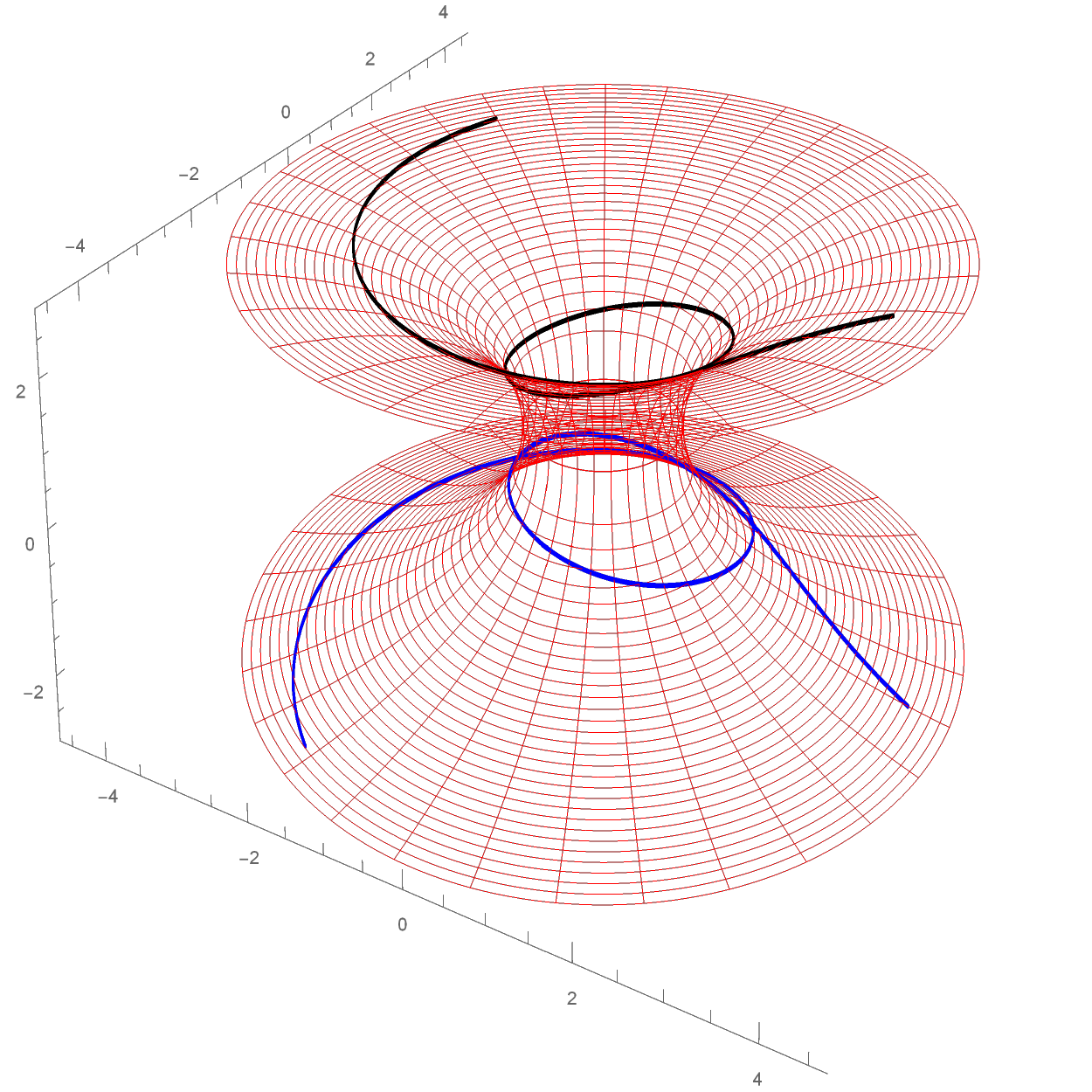}}}\\
	\subfloat[Bound orbit for $ E=25.7, L=15.6, b_0=1 $\label{tidal_bound-d}]{{\includegraphics[height=5cm]{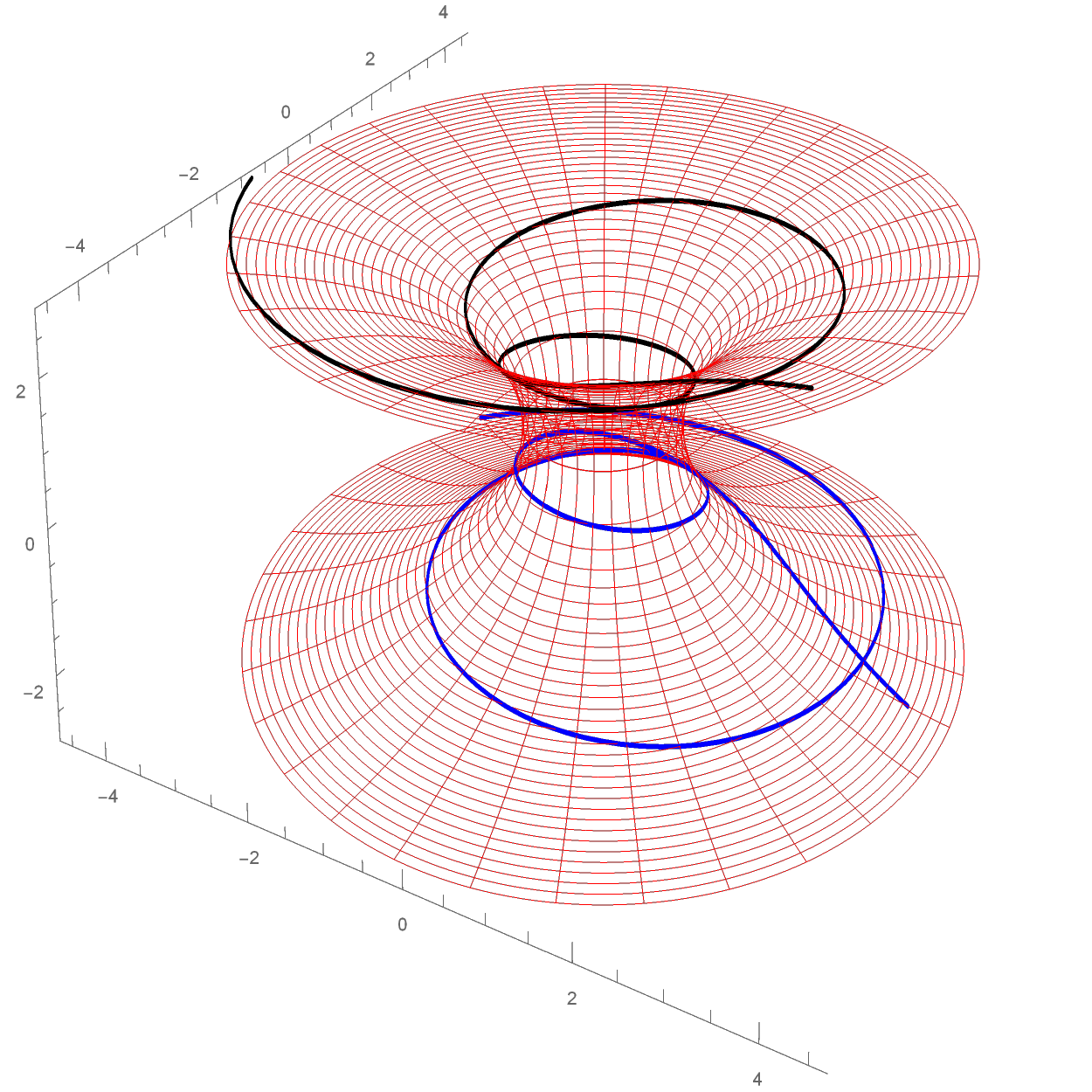}}}\qquad
	\subfloat[Bound orbit for $ E=25.77, L=15.6, b_0=1 $\label{tidal_bound-e}]{{\includegraphics[height=5cm]{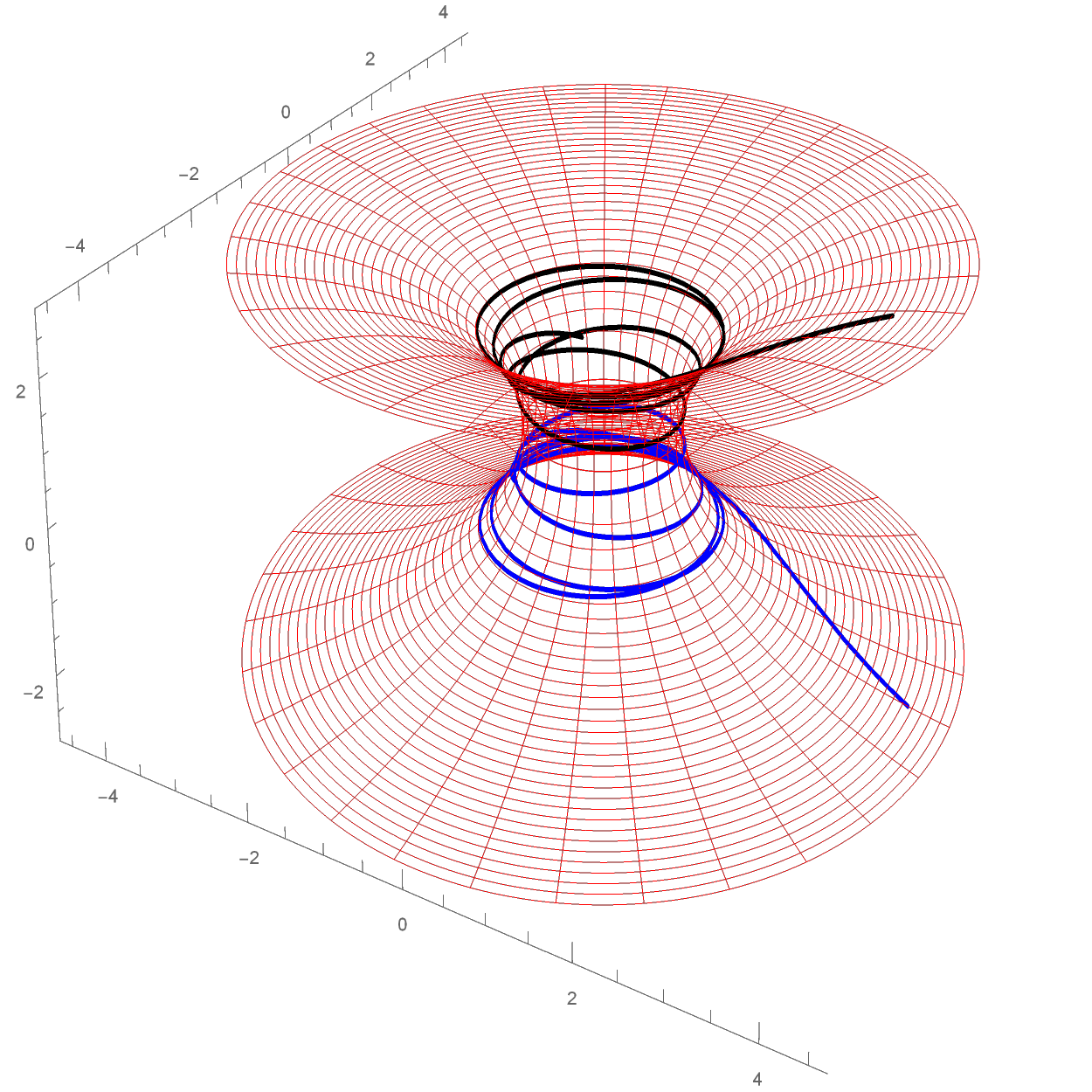}}}\qquad
	\subfloat[Bound orbit for $ E=25.772, L=15.6, b_0=1 $\label{tidal_bound-f}]{{\includegraphics[height=5cm]{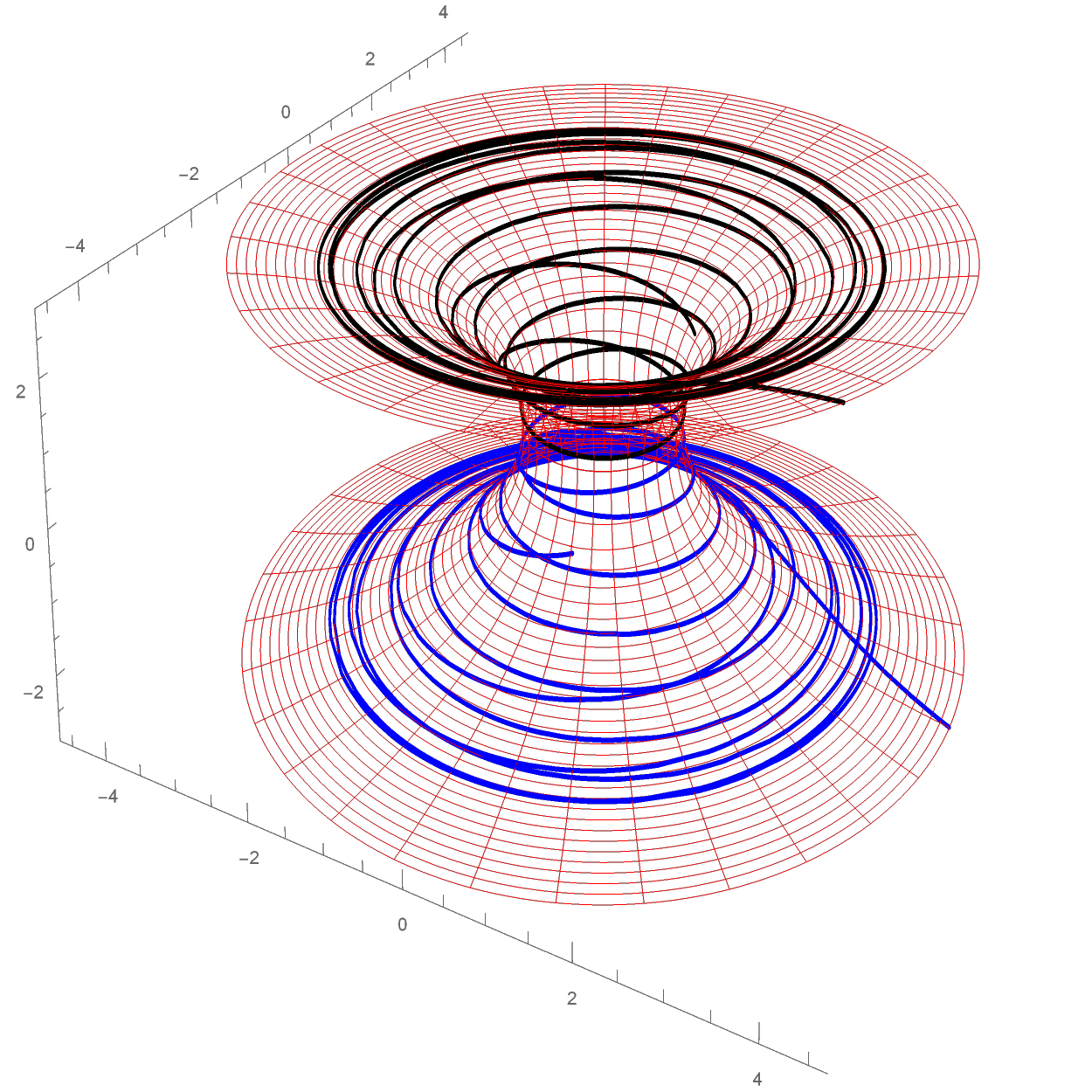}}}
	\caption{Bound orbits for non-zero tidal force (i.e. $ \gamma=1 $) wormhole on the isometric embedding diagram with $ dl/d\phi $ solution for different values of parameters. The throat radius is chosen to be $ b_0=1 $. Variations of the trajectories are exhibited for gradually increasing energy ($ E $) where the angular momentum is fixed at $ L=15.6 $. Orbits started from the upper/lower universe, deflected from the throat region and returned to the same universe. The black and blue trajectories in the upper and lower universe respectively denote the results of positive and negative segments of $ dl/d\phi $. The orbits in \ref{tidal_bound-c}-\ref{tidal_bound-f} contain intersecting closed timelike geodesic trajectories.}
	\label{tidal-bound_plot}
\end{figure*}

\begin{figure}[t!]
	\centerline{\includegraphics[width=5.5cm]{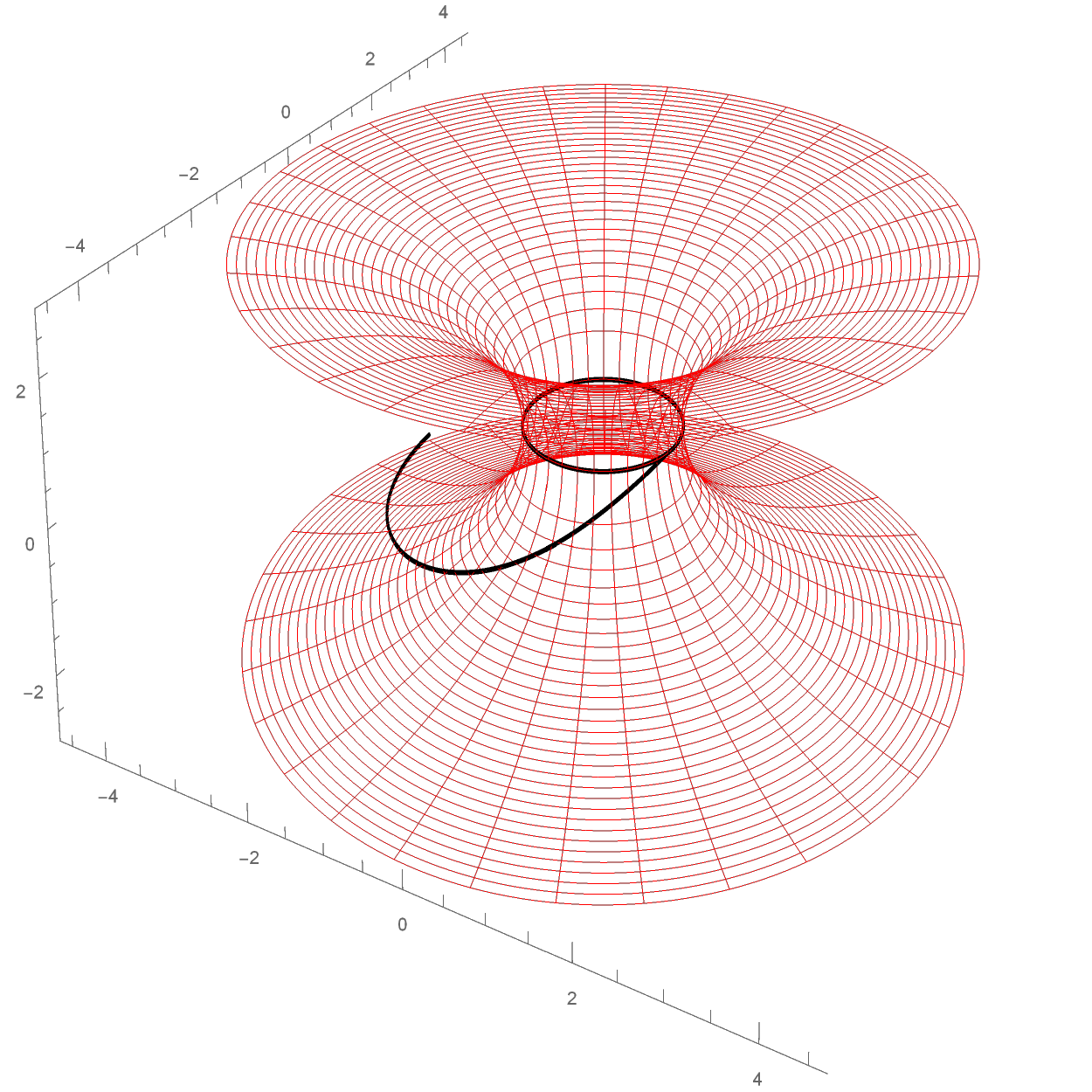}}
	\caption{Bound orbit with $ E = 1545, L = 1566, b_0 = 1 $, containing closed timelike loop. Particles coming from the lower universe enters the closed loop at the throat and stays there for infinite times.}
	\label{throat-CTG_plot}
\end{figure}

To obtain the massive particle orbits, one can write the geodesic equation as
\begin{equation}
	\left(\frac{dr}{d\phi} \right)^2=\left( 1-e^{-(r-b_0)} \right)\left[ \left( \frac{E^2 e^{-\gamma/r}-1}{L^2}\right)r^4-{r^2} \right] .
	\label{orb1}
\end{equation}
where the wormhole shape function and redshift function from Eq. \eqref{shape-eq}, \eqref{red-eq} have been imposed on Eq. \eqref{geo08}. Now, in terms of proper length this equation is modified to
\begin{equation}
	\left(\frac{dl}{d\phi} \right)^2=\left[ \left( \frac{E^2 e^{-\gamma/r(l)}-1}{L^2}\right)r(l)^4-{r(l)^2} \right] .
	\label{orb2}
\end{equation}

Aiming to solve Eq. \eqref{orb2} for its visualization on the wormhole embedding, we recall the equation of embedding function i.e. Eq. \eqref{geo02} and the proper distance given by Eq. \eqref{geo05}. For our choice of shape function (i.e. Eq. \eqref{shape-eq}), the analytical solution of these equations are
\begin{eqnarray}
	z(r) &=& \pm 2 \arctan \sqrt{e^{(r-b_0)}-1} ,
	\label{emb-eq} \\
	l(r) &=& \pm 2 \log[e^{r/2}(1 + \sqrt{1-e^{(b_0-r)}})] .
	\label{orb4}
\end{eqnarray}

Now, in order to plot the trajectories, we invert Eq. \eqref{orb4} to use it in Eq. \eqref{orb2}. As the final expression of Eq. \eqref{orb2} is difficult to solve analytically, the numerical solution of the equation is further used in a 3D parametric plot on wormhole embedding to obtain the orbits. The isometric embedding of the traversable wormhole is visualized by plotting Eq. \eqref{emb-eq}. Finally, the orbits are obtained for different values of energy ($ E $) and angular momentum ($ L $) of the test particles.

The aim of the study is to obtain the bounded and unbounded orbits on the wormhole embedding for zero and non-zero tidal force (i.e. $ \gamma=0~ \text{and}~ \gamma=1 $) wormholes, and thus exploring the possibility of closed timelike geodesics within their trajectories. At the same time, it is necessary to investigate the effect of a tidal force on the geodesics. Thus, various particle motion is exhibited within the wormhole geometry in Figures \ref{escape_plot}, \ref{bound_plot}, \ref{tidal-escape_plot}, and \ref{tidal-bound_plot}. In Fig. \ref{escape_plot} and \ref{bound_plot}, escape and bound orbits of a zero-tidal force wormhole are shown for different values of $ E,~L~ \text{and}~ b_0 $. The evolution of these orbits for non-zero tidal force (with $ \gamma=1 $) are examined in Fig. \ref{tidal-escape_plot}, and \ref{tidal-bound_plot}. For all these trajectories, the throat radius is chose to be $ b_0=1 $, and the angular momentum of the test particle is kept fixed at $ L=15.6 $ to show the variation with increasing energy ($ E $). It is expected to have a smaller deviation in trajectories as the deflection angle is significantly lesser in a wormhole with tidal force. It is however verified in Fig. \ref{tidal-escape_plot}, and \ref{tidal-bound_plot}, where the energy required to mimic a trajectory as in Fig. \ref{escape_plot} and \ref{bound_plot} is higher for the same value of $ L~\text{and}~b_0 $. Further, in bound orbits, particles are subjected to stay in the same universe. Thus, the positive and negative solutions of Eq. \eqref{orb2} draws the same trajectory respectively in the upper and lower universe of the wormhole. These trajectories in lower universe are shown in blue lines as visualized in Fig. \ref{bound_plot} and \ref{tidal-bound_plot}.

It is interesting to note that some of the bound orbits from Fig. \ref{bound_plot}, \ref{tidal-bound_plot} contain intersection points which might be considered as invariant under types of wormhole space-times as previous studies \cite{cataldo2017, taylor2014, willenborg2018} obtained similar results in different circumstances. Referring to \cite{muller2008}, the causal points of light rays have been discussed in detail, where a light flash emitted by an observer at a specific instant of time overlaps with its trajectory at some previous points while traveling on the wormhole embedding. It is the point of interest, if these intersection points called for past events. Notice that at the causal points, a particle has two choices among staying on the present orbit, or taking its past trajectory where after a certain time it travels back to the same point. These intersecting orbits may act like closed timelike or null orbits respectively for massive particles and photons.

In most of the classes of wormholes, intersecting CTG trajectories are seen only in the bound orbits. Thus, a particle necessarily has to be in the same universe (or in the same sublocality of an universe) to attain a CTG motion. However, it may not be impossible to get CTG in escape orbits also, for instance see \cite{willenborg2018}.

Recalling the discussion of section \ref{time-geodesic} involving circular timelike geodesics at the throat, we exhibit a different class of orbit in Fig. \ref{throat-CTG_plot}. However, such orbits have also been obtained previously in \cite{taylor2014, willenborg2018}. It is a perfect example of closed timelike geodesic trajectory at the wormhole throat (in another word, a horizontal slice of the geometry at the throat) which is much similar to the closed timelike orbits of cylindrical wormholes \cite{bronnikov1,bronnikov2,bronnikov3,bronnikov4} and other cylindrical symmetric rotating space-times. The particle coming from the lower universe creates a closed timelike orbit at the throat and rotates infinitely so that it never escapes. In this specific orbit the particle travels for a certain time and returns back to its previous past-self where it entered the loop. A similar kind of circumstances occur in the closed timelike geodesics of cylindrical symmetric space-times as well. Readers are referred to \cite{ijmpd} for a detailed analysis in axially symmetric spacetimes. However, these circular timelike geodesics at the throat are highly unstable as examined by introducing small deviations of parameters.
\\ \\
\textbf{Classification of the $ dl/d\phi $ orbits:}

In the following we shall discuss a set of different trajectories found while the $ dl/d\phi $ motion on the isometric embedding is investigated. The classification of these timelike orbits consisted of the following trajectories
\\
$\bullet$ \textbf{Unbounded escape orbit -} A geodesic trajectory coming from one asymptotic region reaches the other one after approaching and crossing the wormhole throat.
\\
$\bullet$ \textbf{Bound orbit -} A geodesic that effectively remains in the same asymptotic region. In this journey, the geodesic usually deflects from the throat region and returns to the same universe where it came from.
\\
$\bullet$ \textbf{Intersecting CTG orbit -} A geodesic that intersects its trajectory for once or multiple occasions. The overlapping point may introduce the closed timelike geodesic orbits allowing the theoretical time reversal possible. Fig. \ref{bound-c}-\ref{bound-f} and \ref{tidal_bound-c}-\ref{tidal_bound-f} are a few examples of this orbit. Referring to Fig. \ref{bound-f}, \ref{tidal_bound-e}, \ref{tidal_bound-f}, a family of interacting CTGs may also be obtained in this category.
\\
$\bullet$ \textbf{Throat-CTG orbit -} Circular timelike geodesics are only found at the throat which may also attain CTGs as in Fig. \ref{throat-CTG_plot}. However, these trajectories are highly unstable.

\section{Conclusion}\label{conclusion}

In this paper, we intended to study a general wormhole solution in Einstein's gravity by assuming an exponential shape function and a restricted redshift function given by Eq. \eqref{shape-eq} and \eqref{red-eq} respectively, where the $ \gamma $ parameter readily determines the tidal force. We recover ultrastatic wormhole geometry when $ \gamma $ is considered to be zero. On the account of comparing a non-zero tidal force wormhole (with a finite redshift function) with the one with zero-tidal force, $ \gamma=1 $ is considered. In this context, the behaviour of null and weak energy condition components are visualized. Although the energy density is positive as obtained in both the occasions, $ (\rho+p_r) $ is negative at the throat region confirming the presence of exotic fluid for traversability. Subsequently, the need for the exotic matter is found to be higher in finite tidal force wormholes. In summary, the construction of wormhole structure is successfully investigated prior to the discussion of the main aim of the article which is the geodesic motion and the involvement of closed timelike geodesics around these orbits.

In the next section, the geometry of the wormhole is discussed by means of curvature of the spacetime, and the size and shape of the throat. The deflection angle of orbiting photons, as obtained from the geodesic equations, enabled the presence of a repulsive effect of gravity on the geometry. The negative deflection angle which verified the effect is smaller in a non-zero tidal force wormhole than the zero-tidal force, and goes negative for a significantly smaller radial value than the zero-tidal force geometry. However, the nature of the effect is invariant under the choice of shape and redshift function alone. The investigation effectively established that for usual classes of shape functions (such as Ellis wormhole) also provide the property on deflection angle with a non-zero tidal force. Recently, a repulsive gravity effect is found in a numbers of wormhole solutions. It may have a significant role due to the presence of exotic matter at the throat. Thus, the study of this effect in wormhole geometries, and its relation with the type of matter fluid is still an open field of discussion.

It must be noted that both the zero and non-zero tidal force is introduced to investigate the possible effects of tidal force on the particle trajectory and CTG. However, it did indicate that the non-zero tidal force wormhole demands more particle energy to attain the same trajectory and effects as compared to zero-tidal force due to the smaller deflection angle.

In section \ref{time-geodesic}, we examined the circular timelike geodesics which can only be present at the wormhole throat. The bound and unbounded orbits are then discussed in terms of the potential function with proper radial distance. However, the timelike bound orbit condition in a finite redshift as well as in an ultrastatic wormhole are investigated and given by Eq. \eqref{orb_con-tidal} and \eqref{orb_con}. Further, various timelike geodesic orbits parameterized by the energy and angular momentum of the particles, size and shape of the wormhole throat, are presented in a set of figures. Considering the parameters with different values, we construct the geodesic orbits on the isometric wormhole embedding by numerically solving Eq. \eqref{orb2}. In Fig. \ref{escape_plot} and \ref{bound_plot}, escape and bound orbits are respectively exhibited for ultrastatic wormholes where in Fig. \ref{tidal-escape_plot} and \ref{tidal-bound_plot}, those orbits are for the non-zero tidal force wormhole. The gradually increasing energy of the particle ($ E $) for a fixed angular momentum ($ L=15.6 $) shows the respective changes in trajectories for throat radius $ b_0=1 $. The bound orbits that obey Eq. \eqref{orb_con-tidal} and \eqref{orb_con}, and are subjected to get deflected by the throat and consequently stay in the same asymptotic region of the wormhole,have two different solutions for positive and negative segment of Eq. \eqref{orb2}. These orbits mimic each other in the upper and lower universes and are shown in black and blue lines respectively.

Consequently, in this article, we have discussed the existence and nature of closed timelike geodesic trajectories that are effectively found in the bound orbits. In this context, trajectories overlap with its past worldlines while orbiting on the wormhole embedding. In these intersection points, particles are provided with choices to follow the present orbits, or taking the past worldlines where they travel back to the same intersection points after a certain times. This may allow particles to undergo backward time travel within wormhole embedding. Some orbits encounter these causal points once or twice, whereas few undergo a set of (family of) interacting CTGs. The question is whether these CTGs result from the choice of shape function considered in this study (exponential shape function \eqref{shape-eq}). To clarify the doubt, we refer to \cite{cataldo2017,taylor2014,willenborg2018}, where geodesic orbits for a Schwarzschild-like wormhole, asymptotically flat linear factor shape function along with Ellis wormhole shape, and wormhole supported by conformally coupled massless scalar field are discussed in details. Although they did not mention the causal points, all of these studies obtained intersecting orbits, thus allowing for CTGs. Furthermore, it is worth noting that these closed curves are always necessarily geodesics, as it is straightforward to show that these wormholes do not allow the presence of closed timelike curves on the embedding.

On the other hand, the unstable circular timelike orbit at the throat is another source of CTG as obtained in Fig. \ref{throat-CTG_plot}. We have investigated that test particles entering the loop get trapped forever, where they travel back to their past self endlessly after each rotation. Again, referring to \cite{taylor2014}, these orbits are generally possible in wormhole geometries, and is like a copy of what obtained in cylindrical symmetric rotating spacetimes admitting closed timelike orbits, as referred earlier \cite{ijmpd}. At the same time, cylindrical wormholes studied by Bronnikov \textit{et al.} \cite{bronnikov1,bronnikov2,bronnikov3,bronnikov4} discussed closed orbits that may lead to similar investigations.

Finally, we concluded the study by classifying the bounded and unbounded orbits in terms of CTGs present in their trajectories .

\section*{Acknowledgement}
The authors acknowledge the valuable help and support of Akash Bose and Gopal Sardar from SC's lab. S.C. thanks FIST program of DST, Department of Mathematics, JU (SR/FST/MS-II/2021/101(C)).

\end{document}